\documentclass[fleqn,usenatbib]{mnras}
\pdfoutput=1

\usepackage{newtxtext,newtxmath}
\usepackage{graphicx}   % Including figure files
\usepackage{amsmath}    % Advanced maths commands
\usepackage{multirow}
\usepackage[table,xcdraw]{xcolor}
\usepackage{arydshln}

\usepackage[T1]{fontenc}

\DeclareRobustCommand{\VAN}[3]{#2}
\let\VANthebibliography\thebibliography
\def\thebibliography{\DeclareRobustCommand{\VAN}[3]{##3}\VANthebibliography}

\usepackage{graphicx}	% Including figure files
\usepackage{amsmath}	% Advanced maths commands
\usepackage{animate}
\title[Morphology vs. Star Formation]{From Blue Cloud to Red Sequence: Evidence of Morphological Transition Prior to Star Formation Quenching}

\author[Sampaio et al. 2021]{V. M. Sampaio,$^{1}$\thanks{E-mail: vitorms999@gmail.com}
R. R. de Carvalho ,$^{1}$ I. Ferreras,$^{2,3,4}$ A. Arag\'on-Salamanca,$^{5}$ L. C. Parker$^{6}$ 
\\
$^{1}$ NAT - Universidade Cruzeiro do Sul /  Universidade Cidade de S\~ao Paulo, 01506-000, SP, Brazil\\
$^{2}$ Instituto de Astrof\'isica de Canarias, Calle V\'ia L\'actea s/n,
E38205, La Laguna, Tenerife, Spain\\
$^{3}$ Department of Physics and Astronomy, University College London, Gower Street, London WC1E 6BT, UK\\
$^{4}$ Departamento de Astrof\'\i sica, Universidad de La Laguna, E38206 La Laguna, Tenerife, Spain\\
$^{5}$ School of Physics and Astronomy, University of Nottingham, University Park, Nottingham NG7 2RD, UK\\
$^{6}$ Department of Physics and Astronomy, McMaster University, Hamilton ON L8S 4M1, Canada}

\date{Accepted XXX. Received YYY; in original form ZZZ}

\pubyear{2021}
\begin{document}
\label{firstpage}
\pagerange{\pageref{firstpage}--\pageref{lastpage}}
\maketitle

\begin{abstract}
We present a study of a sample of 254 clusters from the SDSS-DR7 Yang Catalog and an auxiliary sample of field galaxies to perform a detailed investigation on how galaxy quenching depends on both environment and galaxy stellar mass. Our samples are restricted to 0.03$\leq$z$\leq$0.1 and we only consider clusters with $\rm log(M_{halo}/M_{\odot}) \geq 14$. Comparing properties of field and cluster galaxies in the Blue Cloud, Green Valley and Red Sequence, we find evidence that field galaxies in the red sequence hosted star formation events $\rm 2.1 \pm 0.7$ Gyr ago, on average, more recently than galaxies in cluster environments. Dissecting the star formation rate vs stellar mass diagram we show how morphology rapidly changes after reaching the green valley region, while the star formation rate keeps decreasing. In addition, we use the relation between location in the projected phase space and infall time to explore the time delay between morphological and specific Star Formation Rate variations. We estimate that the transition from late to early-type morphology happens in $\rm \Delta t_{inf} \sim$1 Gyr, whereas the quenching of star formation takes $\sim$3 Gyr. The time-scale we estimate for morphological transitions is similar to the expected for the delayed-then-rapid quenching model. Therefore, we suggest that the delay phase is characterized mostly by morphological transition, which then contributes morphological quenching as an additional ingredient in galaxy evolution. 
\end{abstract}


%%%%%%%%%%%%%%%%%%%%%%%%%%%%%%%%%%%%%%%%%%%%%%%%%%

%%%%%%%%%%%%%%%%% BODY OF PAPER %%%%%%%%%%%%%%%%%%

\section{Introduction}

Galaxies and their evolution across cosmic time have always intrigued astronomers. In the first half of the twentieth century, Edwin Hubble classified galaxies according to their morphology \citep{1926ApJ....64..321H}, defining two major classes: Early Type (ETGs) -- characterized by elliptical shapes -- and Late Type Galaxies (LTGs) -- which comprise galaxies that are a combination of a central bulge and spiral arms. Later, \cite{Dressler} presents the Morphology-Density (MD) relation, which states that ETGs and LTGs are not uniformly distributed in space. Namely, the former dominates high density environments, while LTGs are mainly found in low density fields. The dependence of galaxy properties with environment was firmly confirmed in the last decades due to the observations provided by wide galaxy surveys such as the Sloan Digital Sky Survey \citep[SDSS]{2000AJ....120.1579Y}, Cosmic Evolution Survey \citep[COSMOS]{2007ApJS..172....1S} and the PRIsm MUlti-Object Survey \citep[PRIMUS]{2011ApJ...741....8C}. A bimodal distribution is found in galaxy color \citep{2020MNRAS.498.6069P} and star formation rate \citep[SFR]{2012MNRAS.424..232W,2020MNRAS.491.5406T} as well. 

We now tend to divide galaxies into three different populations: 1) Blue Cloud (BC) -- filled mainly by late-type (90\% LTG vs 10\% ETG), blue, star forming galaxies; 2) Red Sequence (RS) -- dominated by early-type (70\% ETG vs 30\% LTG), red galaxies with low (if any) star formation; and 3) an intermediate region called the Green Valley (GV) -- containing galaxies partially quenched with intermediate morphologies \citep{2009MNRAS.396..818S, 2010MNRAS.404..792M, 2010MNRAS.405..783M, 2014MNRAS.440..889S}. Additionally, an important caveat is that red LTGs are expected at all masses, whereas blue ETGs are mainly found at the low mass end \citep[e.g.][]{2000ApJ...532..193F,2007ApJS..173..619K}. Finally, a galaxy's path and the related time-scales through the GV depends critically on the quenching mechanism \citep{2014MNRAS.440..889S,2017MNRAS.464.1077W}. 

An important feature is the evolution of cosmic star formation density as a function of time, which has a peak at around $\rm z \sim 2 - 3$ and may influence the relative number density of galaxies in each region depending on the inspected redshift \citep{1996ApJ...460L...1L,1996MNRAS.283.1388M}. This evolution is related to the density fluctuations giving rise to galaxy formation. Namely, galaxies born in early-stage clusters experience high density environments since birth, which is significantly different from the satellite systems infalling in clusters at later stages. The star formation in galaxies is usually fueled by the inflow of hot gas from the circumgalactic medium, which is then cooled via internal interactions in the so-called ``cold-disk'' and eventually gravitationally collapses to form new stars \citep{2019igfe.book.....C}. Quenching then may be related to stopping the hot gas from cooling, for which then the galaxy will passively evolve and only form a negligible amount stars out of the current reservoir of cold gas available in the disk \citep[``slow quenching'']{2015A&A...579A...2I, 2016A&A...590A.103M,2020MNRAS.495.4237M}. When in dense environments, a second way to quenching star formation is to remove all the galaxy's gas component (including the cold disk) in a short time-scale \citep[``rapid-quenching'']{2021essp.confE..59T, 2021essp.confE..39G, 2021ApJ...911...57Z}. 

When a galaxy is isolated, star formation quenching is mostly driven by internal processes \citep{1974MNRAS.169..229L,1986ApJ...303...39D,2021arXiv210402089V}. Active Galatic Nuclei (AGN) feedback create an outflow of gas preventing further hot gas accretion from the circumgalactic medium \citep{2008MNRAS.387.1431D, Bongiorno, 2020MNRAS.491.5406T}. Star formation depends on gravitational instabilities, which may be prevented due to a transition from a stellar disk to a spheroid \citep[``Morphological Quenching'']{2009ApJ...707..250M}. Bars in spiral galaxies may drive gas inflows, which enhance central star formation \citep[Bar-Driven Evolution]{2017MNRAS.465.3729S}. 
However, in the local universe most of the galaxies live in groups/clusters \citep{1983ApJS...52...61G}. Even before crossing the virial radius, infalling galaxies can stop accreting new gas \citep[``Starvation'']{1980ApJ...237..692L,2000ApJ...540..113B,2017MNRAS.466.3460V}. For instance, \cite{2020MNRAS.491.5406T} suggest galaxy quenching has an extended phase ($\sim 5$\,Gyr) of starvation. In addition, infalling galaxies can lose gas, stars and even dark matter via gravitational tides \citep[``Tidal Mass Loss'' - TML]{1999MNRAS.302..771J,2006MNRAS.366..429R}. Within the virial radius, the hot gas in the intracluster medium (ICM) exerts pressure on galaxies moving within the cluster and may remove gas via Ram Pressure Stripping \citep[RPS]{1972ApJ...176....1G,1999MNRAS.308..947A}. Clusters provide a suitable environment for both direct and indirect galaxy interactions, especially in its core. Direct encounters may lead to galaxy mergers and cause a starburst event over a short time scale and quickly exhaust a galaxy's gas supply \citep{2005ApJ...622L...9S,2008MNRAS.384..386C,2010ApJ...720L.149T}, whereas repeated indirect encounters may leave interacting galaxies with disturbed morphologies. At last, it is important to note that clusters are built up by the accretion of galaxy groups. Cluster galaxies may therefore be affected by ``pre-processing'', in which galaxy properties were altered even before entering the cluster \citep{2004PASJ...56...29F,2013MNRAS.431L.117M,2019A&A...632A..49S}.

The balance between internal and environmental processes driving galaxy evolution results in a complex non-linearity. Although \cite{2010ApJ...721..193P} indicate that internal and environmental mechanisms acting on galaxy evolution are separable up to redshift $\rm z \sim 1$, the main mechanism driving galaxy evolution and the related time scales are yet not fully understood. Different works indicate the main quenching mechanism as dependent on galaxy and host halo mass \citep{2016MNRAS.457.4360Z}. \cite{2010ApJ...721..193P} suggest that a stellar mass related mechanism plays a major role in quenching massive galaxies. \cite{2012MNRAS.424..232W} show an increasing fraction of quenched galaxies in clusters for increasing stellar and host halo mass. They also show the quenched fraction grows towards the cluster core. At lower stellar masses, gas outflows are increasingly relevant \citep{2010ApJ...721..193P,2018MNRAS.476.1680S, 2020MNRAS.499..230B}. On sufficiently large time-scales, tidal mass loss can remove a significant fraction of a galaxy's mass \citep[see their Fig.~3]{2017ApJ...843..128R}. Conversely, RPS can remove a significant fraction of gas from low mass galaxies \citep{2016ApJ...826..148E, 2016MNRAS.463.1916F}. Yet, \cite{2019ApJ...873...42R} provide evidence that RPS contributes to the fast quenching ($\sim 2$ Gyr) after galaxy reaches an intra cluster medium density threshold ($\rm log(\rho_{ICM}) \sim -28.5 \, [g \, cm^{-3}]$).

Different models tried to simplify this non-linearity. One of the most successful models is the ``delayed-then-rapid quenching'' \citep{2013MNRAS.432..336W}, in which a galaxy infalling in a cluster is at first unaffected by the high density environment and is mostly quenched due to starvation. After a delay time, environmental effects, especially RPS, rapidly halt galaxy star formation.

An important parameter to understand how galaxies in dense environments transition from star forming to passive/quenched galaxy is the infall time -- the time at which the galaxy has been experiencing the cluster environment. However, it is not possible to directly measure the galaxy's infall time. To estimate the infall time, a usual approach is to invoke the Phase-Space. This is a 6D-space (3 positional and 3 velocity coordinates) widely used to study the dynamics of complex systems. In this space, infalling galaxies have a well defined trajectory, providing a tool to estimate the infall time \citep[see their Fig.~1]{2017ApJ...843..128R}. Galaxies recent infalling have more radial orbits in comparison to those in the virialized state within the cluster, which are characterized mostly by circular orbits \citep{2017ApJ...843..128R}. In observational astronomy we are limited to projected quantities and, consequently, to a projected version of the phase-space: the projected phase-space (PPS). The use of cosmological simulations is pivotal to enable a connection between PPS location and infall time. \cite{2011MNRAS.416.2882M} and \cite{2013MNRAS.431.2307O} show through simulation how different regions of the PPS are mainly occupied by galaxies at different dynamical stages -- and with significantly different orbits -- within the cluster. Using the PPS, \cite{2019MNRAS.484.1702P} show pre-processed galaxies -- those that were already part of a minor group before infalling -- to have different properties from those first experiencing the cluster environment. \cite{2020ApJS..247...45R} combine simulations and observational data to estimate the relation between infall time and SFR. The PPS can be used to constrain regions where different environmental effects are at work. For instance, \cite{2020MNRAS.495..554R} investigate the PPS of the Coma Cluster and find an excess of galaxies possibly suffering RPS in all projected radii within the virial radius at higher velocity offsets. Several works suggest differences between galaxy properties of clusters with different member velocity distributions, which translates to differences in the PPS \citep{2013MNRAS.434..784R, 2017MNRAS.467.3268R, 2019MNRAS.487L..86D, 2021MNRAS.503.3065S}.

The complex relation between infall time, galaxy properties and quenching mechanisms translates to a variety of paths for galaxy's transition from the BC to RS. de Sá-Freitas et al. (submitted) show how different processes lead to different galaxy transition paths (see their Fig.~8) depending on both galaxy properties and the acting mechanism. Consequently, it is not straightforward to reliably define the BC, GV and RS regions, for which different works adopt different parameter spaces and methodologies. For instance, \cite{Strateva1} use the color-color diagram, \cite{2020MNRAS.491.5406T} adopt Stellar Mass versus Star Formation Rate and \cite{2019MNRAS.488L..99A} define quenching according to the spectroscopic measurement of the D4000 break. Although it is known the that morphology correlates with and galaxy properties, it is still necessary to further investigate the transition from BC to RS from a morphological perspective. Also, recent works discuss the possibility of dense environments inducing morphological variations prior to SFR changes \citep{2009ApJ...707..250M,2014MNRAS.440..889S,2019MNRAS.486..868K}.

In this work, we address galaxy evolution from different perspectives. First, we use the SFR vs $\rm M_{stellar}$ plane and the separation between BC, GV and RS to compare cluster and field galaxy properties. We focus on cluster galaxies and investigate their path from BC to RS as a function of stellar mass. We focus on understanding what defined different stages of galaxy evolution and address the question on whether SFR or morphology changes more quickly in cluster environments. Furthermore, we focus on the relation between location in the PPS and infall time to provide direct measurements of how galaxy properties vary with time. Lastly, we investigate how variations in morphology and star formation rate depend on stellar mass and environment.

This paper is organized as follows: in $\S$~2 we present the sample and describe the galaxy properties; in $\S$~3 we compare the overall distributions of cluster and field galaxies; in $\S$~4 we explore the stellar mass vs SFR plane to understand if the usual separation of BC, GV and RS presented in the literature do separate galaxies with respect to stellar population properties. We compare properties of galaxies with different mass and at different environments; in $\S$~5 we define and probe the PPS of galaxy clusters to address variations in morphology and SFR with respect to infall time of member galaxies; in $\S$~6 we discuss our findings; in $\S$~7 we present our final conclusions and a summary. Throughout this paper we adopt a flat $\rm \Lambda CDM$ cosmology with
$[\rm \Omega_{M},\Omega_{\Lambda},H_{0}] = [0.27,0.73,72$ km $\rm s^{-1} Mpc^{-1}]$.

\renewcommand{\arraystretch}{1.5}

\section{Data Selection}
\label{sec:Sample_Data}
In this work, we select galaxies from the Sloan Digital Sky Survey - Seventh Data Release \citep[SDSS-DR7]{2009ApJS..182..543A}. We limit our sample to galaxies within the redshift range $\rm 0.03 \leq z \leq 0.1$ and with petrosian apparent magnitude in the r-band ($\rm m_{r}$) less than 17.78, which correspond to the spectroscopic limit of the survey at z = 0.01. The minimum redshift is applied to avoid biasing the stellar population parameters (see Section \ref{sec:Starlight}) due to the fixed 3 arcsec fiber used in the SDSS. 

\subsection{Galaxy Clusters from the Updated Yang Catalog}
\label{sec:Clusters}

We adopt the Yang Catalog \citep{2007ApJ...671..153Y} to classify galaxies according to their environment. Briefly, the catalog is build by applying a halo mass finder algorithm to the New York University - Value Added Catalog \citep[NYU-VAGC]{2005AJ....129.2562B}. Groups are then defined as galaxies in the same dark matter halo. The Catalog also provides the classification of galaxies into centrals (here we adopt the most massive cluster galaxy as central) and satellites, which enables the definition of: 1) ``isolated centrals'' - defined as halos occupied by a single galaxy; 2) ``cluster centrals'' - which means the most massive galaxy in halos with $\rm N \geq 2$ galaxies; and 3) ``cluster satellites'' - non-central galaxies located at halos with $\rm N \geq 2$ members.

Here we use an updated version of the Yang Catalog presented in \citealt{2017AJ....154...96D} (dC17, hereafter), which is built using data from the SDSS-DR7 with the same redshift range and $\rm m_{r}$ adopted in this work. Unlike in the original Yang Catalog, membership in dC17 is defined via a ``Shiftgapper'' technique (see \citealt{1996ApJ...473..670F} for more details), which is applied in clusters from the Yang Catalog with at least 20 members. Two main reasons favors the use of the Shiftgapper technique to define membership in our case: 1) it avoids prior assumptions about the cluster's dynamical stage; and 2) it is more permissive regarding galaxies at larger distances from the cluster centre, which is relevant for works analysing the evolution of satellite galaxies in higher density environments. Next we briefly describe how this technique works. The first step is to select galaxies at a maximum distance of 2.5h$\rm ^{-1}$ Mpc (3.47 Mpc for h = 0.72) and with a line-of-sight velocity in the range $\rm \pm \, 4000 km\, s^{-1}$ with respect to the clustercentric coordinates (RA, DEC and redshift) presented in the Yang Catalog\footnote{The clustercentric coordinates are the only information from the Yang Catalog used to define the clusters.}. \cite{2017AJ....154...96D} then apply a Gap Technique by placing galaxies in radial bins with a minimum size of 0.42h$\rm ^{-1}$ Mpc, which guarantees at least 15 galaxies per bin, and removing galaxies with a velocity gap greater than $\rm 1000 \, km \, s^{-1}$ with respect to the cluster mean velocity. A new center is then defined as the median RA, DEC and redshift of the remaining galaxies and the process is reiterated until no more galaxies are removed. This process results in a final list containing only member galaxies. 

Using the final list of member galaxies, dynamical quantities like virial radius ($\rm R_{200}$), virial mass ($\rm M_{200}$) and velocity dispersion along the line-of-sight ($\rm \sigma_{LOS}$) are estimated by dC17 through virial analysis (see \citealt{2009MNRAS.399.2201L} and Appendix \ref{AppendixA} for more details) for each cluster. A comparison between the Yang catalog and the shiftgapper technique shows differences in $\rm M_{200}$ of less than 0.1, which are smaller than the related uncertainties in the Yang $\rm M_{200}$ estimates ($\rm \sim 0.15 dex$, \citealt{2007ApJ...671..153Y}). By imposing a minimum number of 20 galaxies within the virial radius\footnote{We define the clustercentric coordinates as the median redshift and the luminosity weighted average of RA and DEC.} we define a sample 319 massive clusters. By using the relation between $\rm M_{200}$ and $\rm N_{200}$, where the last term is the number of galaxies within $\rm R_{200}$, we define a halo mass threshold, namely $\rm M_{200}^{tresh} = 10^{14} \, M_{\odot}$. By considering clusters with $\rm M_{200} \geq M_{200}^{tresh}$, our sample decreases to 254 clusters. The halo mass completeness limit means we are probing clusters at the extreme tail of the halo mass function, namely the 5\% most massive systems with halo masses in the range $\rm 13 \leq log(M_{halo}/M_{\odot}) \leq 15.5$\footnote{We used the Halo Mass Function Calculator \citep{2013A&C.....3...23M} with Planck 15 cosmology \citep{2016A&A...594A..13P} and z = 0.075, which is the median redshift for our sample.}. Additionaly, we make a distinction between satellites and centrals according to the ``rank'' assigned in the Yang Catalog. Here on we focus only on satellite galaxies, which consists of 20,191 galaxies.

\subsection{Sample of Low Interaction Galaxies}
\label{sec:Field_and_Isolated}

Galaxies infalling into clusters are affected even before crossing the system virial radius due to starvation and tidal effects \citep{2020MNRAS.491.5406T}. In addition to our cluster sample, we define a secondary sample of field galaxies to trace how galaxies evolve when in isolation. First, we queried the SDSS-DR7 database for all galaxies with reliable spectroscopic redshift measurements meeting the same selection criteria of our cluster sample. We then select field galaxies as follows: 1) we identify all groups from the Yang Catalog with halo masses greater than $\rm 10^{13} M_{\odot}$ -- a characteristic halo mass of poor groups ($\rm N_{members} \leq 10$); 2) we use the scaling relation 
\begin{equation}
    \rm R_{vir} \sim 1.61 \, Mpc \left ( \frac{M_{halo}}{10^{14}M_{\odot}} \right )^{1/3} \, (1+z_{group})^{-1}
\end{equation}
to estimate the virial radius for every group; 3) we select galaxies beyond 5 $\rm R_{200}$ of every structure with halo mass greater than $\rm 10^{13} M_{\odot}$ listed in the Yang Catalog \citep{2017MNRAS.471L..47T}; 4) we consider galaxies only in the main part of the SDSS footprint ($\rm 120 \leq RA \leq 250$, $\rm 0 \leq DEC \leq 60$) to minimize edge effects. This results in a set of 12,398 galaxies, which is dominated ($\rm \sim 93 \%$) by galaxies classified as Isolated Centrals in the Yang Catalog. To guarantee the reliability of our field sample, we discard all non isolated central galaxies ($\sim 7\%$). Yet, a word of caution is needed regarding fossil groups. These are groups that, after a series of merging events, end up as a single luminous central galaxy. \cite{2009AJ....137.3942L} shows that fossil groups are characterized by a bright galaxy with $\rm M_{r} \leq -22$, where $\rm M_{r}$ is the absolute magnitude in the r-band. Therefore, we adopt this value as the absolute magnitude lower limit for our field sample in order to minimize the effects of fossil groups. This removes $\sim 5\%$ of the our sample and results in a final field sample of 11,674 galaxies, although our results do not change if we keep these galaxies in our sample. This follows from the fact that fossil groups can be considered rare events, which are characterized by a comoving number density of $\rm 2.83 \times 10^{-6} \, h^{3} \, Mpc^{-3}$ \citep{2009AJ....137.3942L}.

\subsection{MPA-JHU Spectral Measurements}
\label{sec:SDSS-DR16}
We retrieve Stellar Mass\footnote{Derived using the methodology of \cite{2003MNRAS.341...33K}} ($\rm M_{stellar}$, hereon) , SFR and sSFR\footnote{Estimates are calculated using \cite{2004MNRAS.351.1151B} prescription and then aperture corrected using the method described in \cite{2007ApJS..173..267S}} estimates from the MPA-JHU catalog, which provides quantities derived from the spectra of SDSS-DR16 galaxies without anomalies in their spectra. We find available estimates for 98 \% of our sample (cluster + field).

The MPA-JHU catalog provide 5 percentiles (2.5, 16, 50, 84 and  97.5\%) for each estimate. Hereafter we use the 50\% percentile (median) as the desired estimate. Regarding the associated uncertainties, we estimate for MPA-JHU quantities from the 16 and 84\% quantiles as $\rm 0.5 \times (Q_{84\%} - Q_{16\%})$. In Table \ref{Table:MPA-JHU_uncertainties}, we present the quantiles, mean and standard deviation for the error distributions. We divide galaxies into three different bins of stellar mass: 1) $\rm 9 \leq log(M_{stellar}/M_{\odot}) < 10$; 2) $\rm 10 \leq log(M_{stellar}/M_{\odot}) < 11$; and 3) $\rm 11 \leq log(M_{stellar}/M_{\odot}) < 12$, following works relating between galaxy properties and stellar mass.

\begin{table}
\caption{Quantiles, mean and standard deviation for the uncertainty distribution for $\rm M_{stellar}$, SFR and sSFR. We present uncertainties for the three stellar mass bins used.}
\label{Table:MPA-JHU_uncertainties}
\resizebox{\columnwidth}{!}{%
\begin{tabular}{c|c|c|c|c|c}
\hline
Uncertainty  & $\rm X = log(M_{stellar})$ & $\rm Q_{16\%}$ & $\rm Q_{50\%}$ & $\rm Q_{84\%}$ & $\rm \mu_{\delta} \pm \sigma_{\delta}$ \\ \hline
\multirow{3}{*}{\begin{tabular}[c]{@{}c@{}}$\rm log(M_{stellar})$\\ $\rm [M_{\odot}]$\end{tabular}} & $\rm 9 \leq X < 10$ & 0.079 & 0.092 & 0.099 & $0.098 \pm 0.058$ \\ \cdashline{2-6}
 & $\rm 10 \leq X < 11$ & 0.088 & 0.092 & 0.098 & $0.099 \pm 0.051$ \\ \cdashline{2-6}
 & $\rm 11 \leq X < 12$ & 0.085 & 0.089 & 0.096 & $0.091 \pm 0.021$ \\ \hline
\multirow{3}{*}{\begin{tabular}[c]{@{}c@{}}$\rm log(SFR)$\\ $\rm [M_{\odot} \, yr^{-1}]$\end{tabular}} & $\rm 9 \leq X < 10$ & 0.222 & 0.326 & 1.009 & $0.507 \pm 0.335$ \\ \cdashline{2-6}
 & $\rm 10 \leq X < 11$ & 0.354 & 1.028 & 1.071 & $0.844 \pm 0.119$ \\ \cdashline{2-6}
 & $\rm 11 \leq X < 12$ & 0.967 & 1.023 & 1.065 & $0.971 \pm 0.178$ \\ \hline
\multirow{3}{*}{\begin{tabular}[c]{@{}c@{}}$\rm log(sSFR)$\\ $\rm [yr^{-1}]$\end{tabular}} & $\rm 9 \leq X < 10$ & 0.242 & 0.342 & 1.013 & $0.523 \pm 0.329$ \\ \cdashline{2-6}
 & $\rm 10 \leq X < 11$ & 0.372 & 1.029 & 1.078 & $0.854 \pm 0.312$ \\ \cdashline{2-6}
 & $\rm 11 \leq X < 12$ & 0.966 & 1.025 & 1.072 & $0.983 \pm 0.174$ \\ \hline
\end{tabular}%
}
\end{table}

\begin{figure}
    \centering
    \includegraphics[width = \columnwidth]{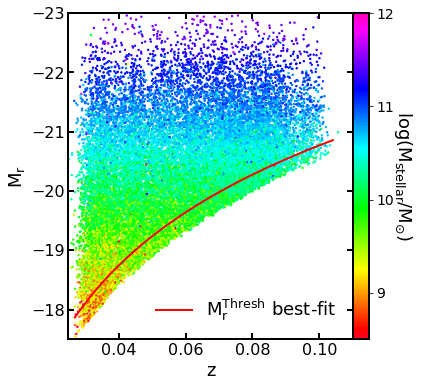}
   \caption{ Relation between absolute magnitude and redshift for our full (Cluster + Field) sample. Points are colored according to galaxy stellar mass. The red line denotes the 90\% $\rm M_{r}$ threshold for each redshift.}
    \label{fig:stellar_mass_thresh}
\end{figure}

We can relate apparent and absolute magnitude at a given redshift. To estimate the absolute magnitude completeness limit in our sample, we: 1) plot absolute magnitude ($\rm M_{r}$) as a function of redshift; 2) divide galaxies into bins of size 0.005 in $\rm z$ and ranging from $\rm 0.03 \leq z \leq 0.1$; 3) define for each bin the 90\% percentile of the $\rm M_{r}$ distribution; 4) apply a logarithmic fit\footnote{We make use of the statistical analysis tool SciDavis \citep{benkert2014scidavis}} (taking into account errors in both x and y axis); and 5) for each redshift we have a limiting absolute magnitude. In Fig.~\ref{fig:stellar_mass_thresh} we present this procedure, which ensures completeness in $\rm M_{stellar}$. However, this process means that our sample is restricted to decreasing redshift for decreasing stellar mass. For instance, only systems with $\rm M_{stellar} \geq 10^{10.7} M_{\odot}$ are considered in the entire redshift of our sample.

\subsection{Morphological Characterization}
\label{sec:Morphological_Params}

An important feature of galaxy evolution is the morphological transition from late to early-type morphology. To explore this transition, we use the TType parameter to trace morphology. This parameter was first introduced by \cite{1963ApJS....8...31D} in order to classify lenticular (S0) galaxies (TType = 0). $\rm TType < 0$ denotes early-type, whereas $\rm TType > 0$ represents late-type galaxies. We select TType estimates from \cite{2018MNRAS.476.3661D} catalog, which uses Convolutional Neural Network based Deep Learning Algorithms to classify the TType of 670,722 galaxies from the SDSS database. We highlight that, differently from the original discrete TType definition, in this case it is measured as a real number. 

Assessing the uncertainty in TType values is not straightforward.  By using the Galaxy Zoo 2 \citep{2013MNRAS.435.2835W} questions and classification for  240,000 galaxies with $\rm m_{r} < 17$ and z < 0.25, \cite{2018MNRAS.476.3661D} show that their method is able to recover 97\% of the ``true'' classification, independent of redshift and apparent magnitude. Yet, an important caveat is the mixture between Elliptical and Lenticular galaxies (see their Fig.~13). To address this separation, they use a more focused model for galaxies with TType $\leq$ 0. In this case, their model is able to separate Elliptical and Lenticular galaxies with an 86\% confidence level. The reliability of the TType estimates is explored in \cite{2020A&C....3000334B}, who finds an agreement between machine learning classification between spiral and elliptical morphologies (see their Figs. 11 and 12 for comparison) and the values provided by \cite{2018MNRAS.476.3661D}.

\subsection{Spectral Fitting Derived Parameters}
\label{sec:Starlight}

Galaxy evolution is directly related to stellar population properties. As time passes, a galaxy's stellar component becomes older and more metal-rich. However, these quantities can not be directly measured and hence are inferred from galaxy spectra. Here we select age and Stellar Metallicity (simply [Z/H] hereon) from the dC17 catalog, which provides stellar parameter estimates for 570,685 galaxies with reliable\footnote{Selected via $\rm zWarning = 0$.} spectra in the SDSS-DR7 database. Briefly, it uses the STARLIGHT code \citep{2005MNRAS.358..363C} to perform a linear combination of predefined single stellar population (SSP) in order to get the best fit to the observed spectra. The fit uses the Medium resolution INT Library of Empirical Spectra \citep{2006MNRAS.371..703S} as stellar model, which is characterized by an almost constant spectral resolution of $\sim 2.5$\,\AA. The SSP grid is built with steps of $0.2$ dex in $\rm log(age)$, ranging from 0.07 to 14.2 Gyr, and includes five possible metallicities: \{-1.71, -0.71, -0.38, 0.00, +0.20\} (see Section 2 of dC17 for more details). We use the luminosity weighted derived age and [Z/H], where the former is closely related to the last star formation episode \citep[T20, hereafter]{2020MNRAS.491.5406T}.

\begin{table}
\centering
\caption{Median and standard deviation of the uncertainty distribution in age and [Z/H] derived using the STARLIGHT spectral fitting method alongside with a set of repeated observation from the SDSS galaxy database. We separate galaxies according to their stellar mass.}
\label{table:starlight_errors}
\resizebox{0.85\columnwidth}{!}{\begin{minipage}{\columnwidth}
\begin{tabular}{c|c|c}
\hline
$\rm X = log(M_{stellar})$ & $\Delta$Age {[}Gyr{]} & $\Delta${[}Z/H{]} \\ \hline
$\rm 9 \leq X < 10$ & $0.32 \pm 0.62$ & $0.005 \pm 0.024$ \\
$\rm 10 \leq X < 11$ & $0.27 \pm 0.33$ & $0.007 \pm 0.019$ \\
$\rm 11 \leq X < 12$ & $0.23 \pm 0.42$ & $0.006 \pm 0.032$ \\ \hline
\end{tabular}%
\end{minipage}}
\end{table}

\begin{figure*}
    \centering
    \includegraphics[width = \textwidth]{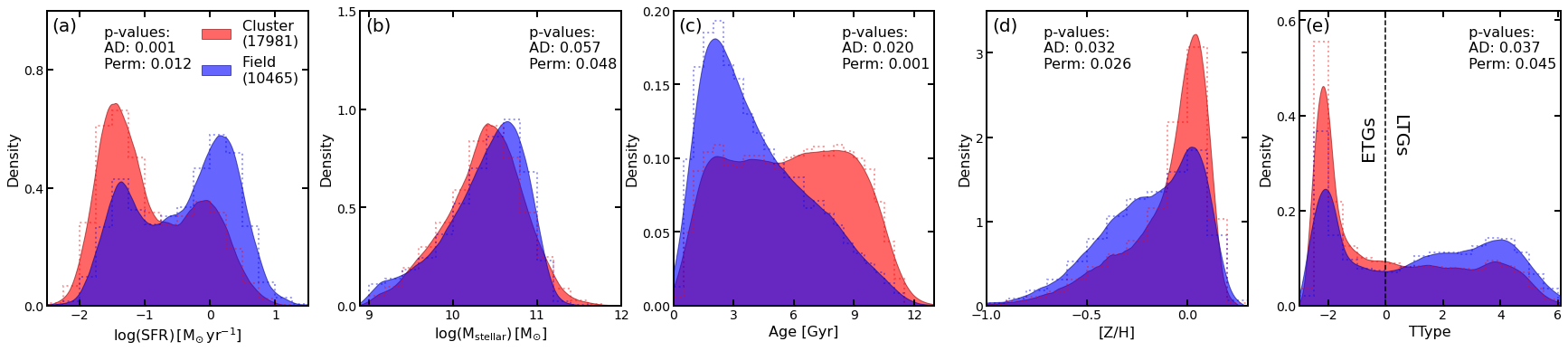}
    \caption{Distribution of SFR, $\rm M_{stellar}$, age, [Z/H] and TType. We separate galaxies according to their environment into clusters (red) and field (blue). The dashed lines represent the distributions histogram, while the filled area represent an epanechnikov kernel density estimate with bandwidth set equal to half of the dashed line histogram bins. The dark dashed vertical line in panel [e] denote the separation between early and late type morphology. We also add in each panel the resulting p-value of Anderson-Darling (AD) and Permutation (Perm) statistical tests. P-values smaller than 0.05 means the two distributions are statistically different.}
    \label{fig:Full_Properties}
\end{figure*}

The estimates of stellar population parameters have an uncertainty due to the observed spectra, which can vary over different observations. We then use a set of galaxies with repeated observations in the SDSS-DR7 covering the same redshift range and magnitude limit of our Samples to assess the stellar population parameter uncertainties. We further impose a minimum signal to noise ratio of 20 in the observed spectra. This selection criteria results in 2,543 galaxies summing up to 6,148 observations. We use the direct comparison of same galaxy observations to derive the expected uncertainties, which are presented in Table \ref{table:starlight_errors} for three stellar mass bins.

\section{Environmental Impact on the Observed Bimodality}

The observed bimodality in certain galaxy properties has been shown to depend on stellar mass. In particular, the bimodality is stronger at lower stellar masses. In this section we focus on exploring how the bimodality in different galaxy properties depend not only on stellar mass, but on environment as well. In Fig.~\ref{fig:Full_Properties} we show the distributions for SFR, $\rm M_{stellar}$, age, [Z/H] and morphology (TType) for galaxies in clusters (red) and in the field\footnote{Low interacting galaxies} (blue). The distributions are build using a Epanechnikov kernel density estimation with bandwidth set to 1.5 times the bin of the dotted histograms in Fig.~\ref{fig:Full_Properties}. We apply the kernel density estimate techniques directly on the data and only use the histograms for the choice of bandwidth. We statistically compare the distributions using 2-sample hypothesis tests. In each panel we display the resulting p-value of k-sample Anderson-Darling and Permutation tests. We decide to use two different statistical tests so that our results are free of any underlying hypothesis of such tests. We select Anderson-Darling and permutation tests due to the different approach they use to measure similarities between the distributions, which enhance the reliability in the derived p-values and distribution comparison. We adopt $\alpha = 0.05$ as the significance level.

Our results reinforce previously found trends. Properties of cluster and field galaxy are significantly different, which is statistically confirmed by the p-values shown in each panel. In panel [a], we find an excess of quenched galaxies (log(SFR) < -1 ) in clusters, whereas field galaxies have more active star formation, which can be seen by the excess of field galaxies with log(SFR) > -0.5. Note that the stellar mass distributions are marginally similar (see panel [b]), which show that the differences in star formation history and environment result in different stellar population properties, at fixed stellar mass. These differences are also seen in Age (panel [c]), in which field galaxies have had a more recent star formation episode\footnote{Age here is closely related to the last star formation episode.} (blue peak at lower ages) in comparison to clusters. These trends indicate how environment is directly affecting the observed bimodality and is further reinforced exploring [Z/H] (panel [d]). Older galaxies are expected to have higher metallicities, for which we see an excess of more metal-poor galaxies in the field in comparison to clusters. Ultimately, environment affects morphology and we find a percentage excess of LTGs in the field in comparison to clusters, in agreement with the Morphology-Density relation.

In this section we investigate overall comparisons between galaxy properties inhabiting distinct environments. The reported results indicate how environment plays a major role in galaxy evolution even when considering global distributions such as Fig.~\ref{fig:Full_Properties}. Our results suggest that galaxies in the field hosted more recent star formation episodes in comparison to those in clusters. We call attention to the presence of a significant fraction of quenched galaxies in the field. This shows how quenching mechanisms unrelated to environment are sufficient to quench field galaxies, whilst environmental effects act like a catalyst for the quenching process, which happens faster in clusters. Namely, in dense environments: 1) galaxy formation can happen earlier due to larger density fluctuations; and 2) clusters add new channels for star formation quenching through interactions, which are not seen for galaxies evolving in isolation. The next step is to consider not only global distributions, but divide galaxies into BC, GV and RS subsamples.

\section{Tracing A Galaxy Path Towards the Red Sequence}

Galaxy evolution from the BC to the RS is directly related to the quenching of star formation. However, galaxy evolution is seen in other properties as well. Quiescent galaxies are mostly early-type \footnote{Note that 30\% of RS galaxies are LTGs and 10\% of BC galaxies are ETGs.}, older and more metal-rich than star-forming ones. It is hence important to understand whether popular definitions of the green valley cover all the properties expected for a transitioning galaxy. A common GV definition is based on the $\rm log(SFR) \, vs \, log(M_{stellar})$ plane, which traces mostly the current star formation, whereas the D4000 break definition traces the current stellar population. We adopt the definition presented in T20, which use the relations
\begin{equation}
    \rm log(SFR) = 0.7 log(M_{stellar}) - 7.52
\end{equation}
to divide BC and GV galaxies and 
\begin{equation}
    \rm log(SFR) = 0.7 log(M_{stellar}) - 8.02
\end{equation}
to separate GV and RS. Further, galaxy evolution is known to be mass-dependent. We hence divide galaxies into three logarithmic stellar mass bins ($\rm log(M_{stellar}/M_{\odot})$ of 9 to 10, 10 to 11 and 11 to 12), which are selected to trace different star formation histories \citep{2012ApJ...752L..27T}. In Table \ref{tab:galaxies_in_each} we display the number of Cluster and Field galaxies in each stellar mass regime and location in the $\rm log(SFR) \, vs. \, log(M_{stellar})$ plane. In a few words, the RS regions becomes more populated for increasing stellar mass and environment density. 
\begin{table}
\centering
\caption{The number (percentage) of cluster or field galaxies in a given stellar mass range and location in the log(SFR) vs log($\rm M_{stellar}$) plane. BC, GV and RS are defined following T20}
\label{tab:galaxies_in_each}
\resizebox{0.85\columnwidth}{!}{%
\begin{tabular}{cc|c|c|c}
\cline{3-5}
 &  & \multicolumn{3}{c}{Region in Plane} \\ \hline
\multicolumn{1}{c|}{Sample} & $\rm X = log(M_{stellar})$ & BC (\%) & GV (\%) & RS (\%) \\ \hline
\multicolumn{1}{c|}{\multirow{3}{*}{Cluster}} & $\rm 9 \leq X < 10$ & 1727 (61\%) & 330 (11\%) & 816 (28\%) \\
\multicolumn{1}{c|}{} & $\rm 10 \leq X < 11$ & 1991 (21\%) & 970 (10\%) & 6519 (69\%) \\
\multicolumn{1}{c|}{} & $\rm 11 \leq X < 12$ & 55 (6\%) & 57 (6\%) & 855 (88\%) \\ \hline
\multicolumn{1}{c|}{\multirow{3}{*}{Field}} & $\rm 9 \leq X < 10$ & 2074 (87\%) & 122 (5\%) & 185 (8\%) \\
\multicolumn{1}{c|}{} & $\rm 10 \leq X < 11$ & 4064 (47\%) & 916 (11\%) & 3667 (42\%) \\
\multicolumn{1}{c|}{} & $\rm 11 \leq X < 12$ & 93 (13\%) & 93 (13\%) & 460 (74\%) \\ \hline
\end{tabular}
}
\end{table}

\begin{figure}
    \centering
    \includegraphics[width = \columnwidth]{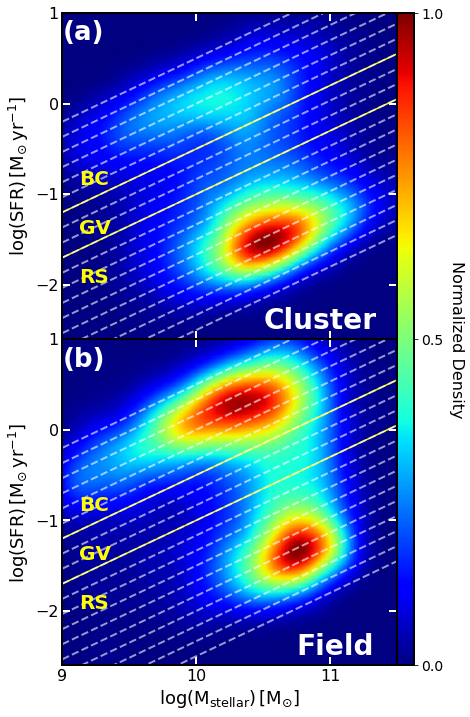}
   \caption{Kernel smoothed normalized density in the SFR vs $\rm M_{stellar}$ plane for cluster (panel [a]) and field (panel [b]) galaxies. Yellow lines denote the limits of BC, GV, and RS according to T20. The white dashed lines have slope equal to equation (1), but with varying intercept from 6.3 to 9.4 in steps of 0.17 (see Section \ref{sec:TTRD} for an explanation).}
    \label{fig:SFR_Mstellar_density}
\end{figure}

In Fig.~\ref{fig:SFR_Mstellar_density} we present an adaptive kernel density estimate for the cluster (panel [a]) and field (panel [b]) galaxy SFR vs $\rm M_{stellar}$ distributions. The limiting lines between BC/GV and GV/RS are shown in yellow. In panel [a], we find a single high density peak at the RS, which comprises $\sim$ 62\% of the cluster sample. This indicates that more than half of the cluster galaxies reached a passive state, where most of evolution is driven by stellar evolution. However, it is important to stress that galaxies can also experience an ``inverse evolution'', in which galaxies go from the RS to the BC due to rejuvenation processes \citep[e.g.][]{2016MNRAS.460.3925T}. Cluster Galaxies in the RS are characterized by median values of $10.73 \pm  0.23$ and $-1.64 \pm  0.19$ in $\rm log(M_{stellar})$ and log(SFR), respectively. On the other hand, the bimodality is more striking in the field sample. Panel [b] shows how most of the star forming galaxies are found in the field, namely $\sim 53\%$ and $\sim 38\%$ of field galaxies are in the BC and RS, respectively. It is clear that most of the active star formation is happening in galaxies with lower stellar mass, which is in agreement with the downsizing scenario, in which most of the star formation in the current universe is happening in low mass galaxies \citep{2006MNRAS.372..933N}.

\begin{figure*}
    \centering
    \includegraphics[width = \textwidth]{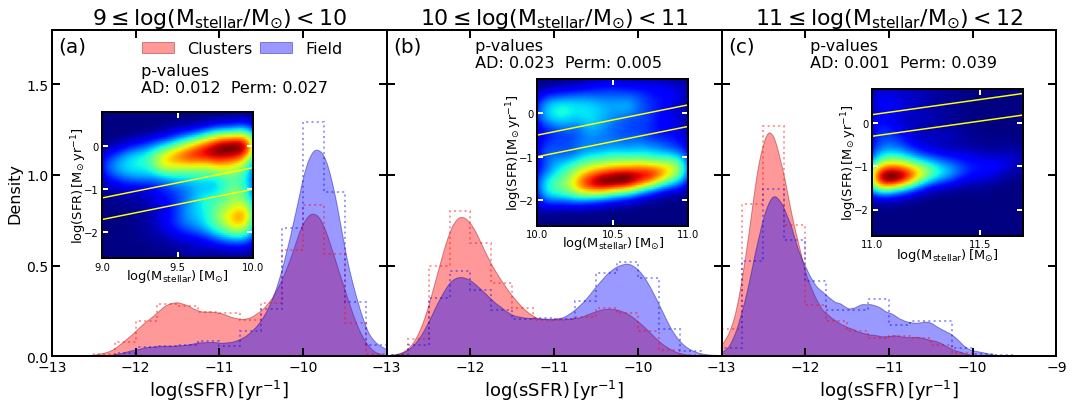}
   \caption{Distributions of sSFR for cluster (red) and field (blue) galaxies for the low (a), intermediate (b) and high (c) stellar mass regimes. We add the mass range in the top of each panel. The filled area is an epanechnikov kernel density estimate using bandwidth equal to 2 times the bin of the displayed histogram in dashed lines. In each panel we present the resulting p-values of a permutation (Perm) and Anderson-Darling (AD) statistical test. For comparison, we include an inset of the relevant part of Fig's.~\ref{fig:SFR_Mstellar_density} panel [a] (cluster galaxies) for each panel.}
    \label{fig:sSFR_hist}
\end{figure*}

A fundamental parameter in astrophysics is the specific SFR (sSFR), which is defined as the SFR divided by the galaxy's stellar mass. This quantity is usually reported in $yr^{-1}$ units, hence its inverse gives an estimate of a time-scale the galaxy would take to form its stellar component. In other words, it would take the inverse of the galaxy's sSFR for it to form all its stars at the current SFR. In Fig.~\ref{fig:sSFR_hist} we present the distribution of log(sSFR) for cluster (red) and field (blue) galaxies for each stellar mass bin. The shaded areas are built using an epanechnikov kernel density estimator with bin width equal to 0.25 in both cases. For comparison, we add a miniature version of the related SFR vs $\rm M_{stellar}$ plane for cluster galaxies in each panel. We compare the distributions using the permutation (perm) and Anderson-Darling (AD) two sample statistical tests and report the p-values in each panel. Exploring the low mass regime (panel [a]), we find $\sim$84\% galaxies with log(sSFR) > -10.5, in comparison to $\sim$62\% for cluster galaxies. We find an excess of low sSFR galaxies in clusters in comparison to the field (26\% and 9\%, respectively, have log(sSFR)<-11). It is important to stress that, even in clusters, low mass galaxies are those predominantly in the BC as can be seen by the inset in panel [a]. In panel [b] and [c], we see an increase of quenched fraction with increasing stellar mass. Panel [b] displays the distribution peaks at different log(sSFR), for which we find cluster galaxies to have dominantly low sSFR in comparison to the peak we find for the field. Finally, in panel [c], massive galaxies are predominantly quenched, independent of environment. This suggests that massive galaxies can rely mostly on internal mechanisms to halt their star formation and do not strongly depend on environmental effects as is the case for lower mass galaxies. Yet we do find statistically different distributions for cluster and field massive galaxies, which may indicate the major effect of environmental is to simply ``accelerate the proccess''.

\subsection{Cluster vs. Field Galaxy Properties}

\begin{figure*}
    \centering
    \includegraphics[width = \textwidth]{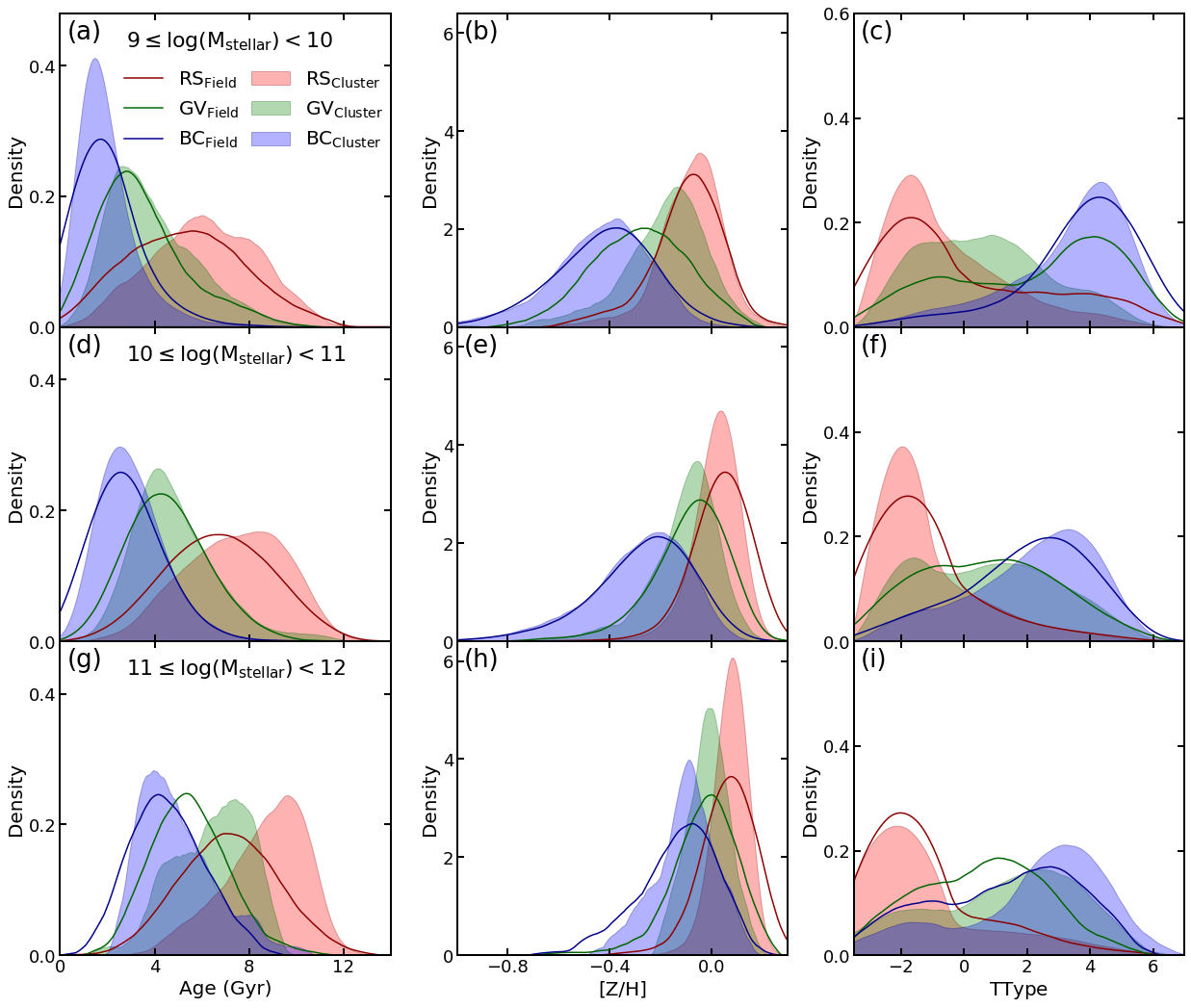}
    \caption{Each panel is an epanechnikov kernel density estimate for the distributions of of age, [Z/H] and TType for galaxies in RS (red), GV (green) and BC (blue). Cluster galaxy distributions are shown as filled areas, whereas solid lines  represent field distributions. Each row corresponds to a stellar mass bin, which is presented in the first column.}
    \label{fig:Cluster_vs_Field_properties}
\end{figure*}

Galaxies in different environments are affected by different quenching mechanisms, which leave an imprint on galaxy properties. In Fig.~\ref{fig:Cluster_vs_Field_properties} we present the distributions of age, [Z/H] and TType of cluster and field galaxies. We separate them according to stellar mass regime and location in the SFR vs. $\rm M_{stellar}$ diagram (RS - red, GV - green and BC - blue). The filled areas represent the distribution for clusters, whereas solid lines denote its field counterpart. In Table \ref{tab:p_values_field_cluster} we show the resulting p-values for AD and Permutation tests comparing cluster and field galaxy age, [Z/H] and TType distributions. The highlighted cells correspond to distributions statistically different according to AD and permutation tests.

\begin{table}
\caption{The resulting p-values of AD and Permutation tests comparing the distributions of cluster and field galaxy ages, [Z/H] and TTypes. We separate galaxies according to luminosity and location in the SFR vs. $\rm M_{stellar}$ plane. We highlight in red the statistically different distributions.}
\label{tab:p_values_field_cluster}
\resizebox{\columnwidth}{!}{%
\begin{tabular}{ccccccccccc}
\cline{3-11}
 &  & \multicolumn{9}{c}{$\rm X = log(M_{stellar})$} \\ \cline{3-11} 
 &  & \multicolumn{3}{c|}{$\rm 9 \leq X < 10$} & \multicolumn{3}{c|}{$\rm 10 \leq X < 11$} & \multicolumn{3}{c|}{$\rm 11 \leq X < 12$} \\ \cline{3-11} 
 &  & \multicolumn{1}{c|}{Age} & \multicolumn{1}{c|}{{[}Z/H{]}} & \multicolumn{1}{c|}{TType} & \multicolumn{1}{c|}{Age} & \multicolumn{1}{c|}{{[}Z/H{]}} & \multicolumn{1}{c|}{TType} & \multicolumn{1}{c|}{Age} & \multicolumn{1}{c|}{{[}Z/H{]}} & TType \\ \hline
\multicolumn{1}{c|}{} & \multicolumn{1}{c|}{Perm} & \multicolumn{1}{c|}{0.204} & \multicolumn{1}{c|}{0.069} & \multicolumn{1}{c|}{0.073} & \multicolumn{1}{c|}{0.234} & \multicolumn{1}{c|}{0.403} & \multicolumn{1}{c|}{\cellcolor[HTML]{FFCCC9}0.012} & \multicolumn{1}{c|}{\cellcolor[HTML]{FFCCC9}0.003} & \multicolumn{1}{c|}{0.792} & 0.106 \\ \cdashline{2-11}
\multicolumn{1}{c|}{\multirow{-2}{*}{\begin{tabular}[c]{@{}c@{}}Blue\\ Cloud\end{tabular}}} & \multicolumn{1}{c|}{AD} & \multicolumn{1}{c|}{0.362} & \multicolumn{1}{c|}{0.088} & \multicolumn{1}{c|}{0.053} & \multicolumn{1}{c|}{0.435} & \multicolumn{1}{c|}{0.672} & \multicolumn{1}{c|}{\cellcolor[HTML]{FFCCC9}0.028} & \multicolumn{1}{c|}{\cellcolor[HTML]{FFCCC9}0.009} & \multicolumn{1}{c|}{0.603} & 0.163 \\ \hline
\multicolumn{1}{c|}{} & \multicolumn{1}{c|}{Perm} & \multicolumn{1}{c|}{\cellcolor[HTML]{FFCCC9}0.021} & \multicolumn{1}{c|}{\cellcolor[HTML]{FFCCC9}0.001} & \multicolumn{1}{c|}{\cellcolor[HTML]{FFCCC9}0.001} & \multicolumn{1}{c|}{0.304} & \multicolumn{1}{c|}{\cellcolor[HTML]{FFCCC9}0.015} & \multicolumn{1}{c|}{\cellcolor[HTML]{FFCCC9}0.013} & \multicolumn{1}{c|}{\cellcolor[HTML]{FFCCC9}0.006} & \multicolumn{1}{c|}{0.842} & \cellcolor[HTML]{FFCCC9}0.039 \\ \cdashline{2-11}
\multicolumn{1}{c|}{\multirow{-2}{*}{\begin{tabular}[c]{@{}c@{}}Green\\ Valley\end{tabular}}} & \multicolumn{1}{c|}{AD} & \multicolumn{1}{c|}{\cellcolor[HTML]{FFCCC9}0.018} & \multicolumn{1}{c|}{\cellcolor[HTML]{FFCCC9}0.002} & \multicolumn{1}{c|}{\cellcolor[HTML]{FFCCC9}0.001} & \multicolumn{1}{c|}{0.607} & \multicolumn{1}{c|}{\cellcolor[HTML]{FFCCC9}0.004} & \multicolumn{1}{c|}{\cellcolor[HTML]{FFCCC9}0.019} & \multicolumn{1}{c|}{\cellcolor[HTML]{FFCCC9}0.002} & \multicolumn{1}{c|}{0.352} & \cellcolor[HTML]{FFCCC9}0.018 \\ \hline
\multicolumn{1}{c|}{} & \multicolumn{1}{c|}{Perm} & \multicolumn{1}{c|}{\cellcolor[HTML]{FFCCC9}0.002} & \multicolumn{1}{c|}{0.386} & \multicolumn{1}{c|}{\cellcolor[HTML]{FFCCC9}0.009} & \multicolumn{1}{c|}{\cellcolor[HTML]{FFCCC9}0.001} & \multicolumn{1}{c|}{\cellcolor[HTML]{FFCCC9}0.003} & \multicolumn{1}{c|}{0.447} & \multicolumn{1}{c|}{\cellcolor[HTML]{FFCCC9}0.001} & \multicolumn{1}{c|}{0.053} & \cellcolor[HTML]{FFCCC9}0.001 \\ \cdashline{2-11}
\multicolumn{1}{c|}{\multirow{-2}{*}{\begin{tabular}[c]{@{}c@{}}Red\\ Sequence\end{tabular}}} & \multicolumn{1}{c|}{AD} & \multicolumn{1}{c|}{\cellcolor[HTML]{FFCCC9}0.003} & \multicolumn{1}{c|}{0.319} & \multicolumn{1}{c|}{\cellcolor[HTML]{FFCCC9}0.002} & \multicolumn{1}{c|}{\cellcolor[HTML]{FFCCC9}0.001} & \multicolumn{1}{c|}{\cellcolor[HTML]{FFCCC9}0.001} & \multicolumn{1}{c|}{0.274} & \multicolumn{1}{c|}{\cellcolor[HTML]{FFCCC9}0.001} & \multicolumn{1}{c|}{0.067} & \cellcolor[HTML]{FFCCC9}0.001 \\ \hline
\end{tabular}%
}
\end{table}

Exploring Fig.~\ref{fig:Cluster_vs_Field_properties}, we find increasing differences between cluster and field distributions with increasing stellar mass. Regarding age (first column panels), we find the most striking differences in the GV and RS distributions, in which field galaxies have an excess of lower age galaxies in comparison to clusters. For instance, 50.58\% of the most massive cluster galaxies in the RS have age > 8 Gyr, while this percentage decreases to 28.06\% for field galaxies. In other words, this excess of younger GV/RS galaxies in the field suggests more recent star formation episodes in these systems in comparison to their counterpart in clusters. On the other hand, the differences we find in [Z/H] (central column panels) are nuanced in comparison to those in age. Qualitatively, field GV/RS [Z/H] distributions have a slight excess of more metal-rich systems in comparison to clusters. However, the [Z/H] distribution for the most massive GV galaxies in the field extends to lower metallicities in comparison to clusters. This is in agreement with previous works showing differences in spectral lines and structure of elliptical galaxies in clusters in comparison to those in the field \citep{1999ApJ...527...54B,2017A&A...597A.122S}. Regarding morphology, panel [c] indicates an excess of GV galaxies with lower values of TType in clusters, when compared to the field, which is further evidence of the environment acting on low-mass members.

\subsection{Towards the Red Sequence}
\label{sec:TTRD}
In the last section we detail, for a given galaxy environment and location in the SFR vs. $\rm M_{stellar}$ plane, significant differences in the age, [Z/H] and TType distributions between field and cluster galaxies. In Fig.~\ref{fig:Cluster_vs_Field_properties}, the intersections between the galaxy property distributions in the BC, GV and RS indicate that the definition from the SFR vs $\rm M_{stellar}$ perspective alone is insufficient to categorize galaxies. An intrinsic problem is that galaxy evolution is a continuous process, while the usual approach is to discretize galaxy populations to just three regions. In order to further understand how galaxies evolve from the BC towards the RS, we create 17 different regions in the SFR vs $\rm M_{stellar}$ plane for the cluster and field sample. The limits of each region are defined as:
\begin{equation}
    \rm log(SFR) = 0.7 log(M_{stellar}) - i
\end{equation}
with i varying from 6.5 to 9.5 in steps of 0.17, which guarantees a minimum of $\sim 50$ galaxies per slice. We present these regions as white dashed lines in Fig.~\ref{fig:SFR_Mstellar_density}. We then select galaxies within each slice and create normalized kernel density estimates for age, [Z/H] and TType. An important feature is the different number of galaxies in each slice. We then define a number-dependent bandwidth for the kernel density estimate, given as:
\begin{equation}
    \label{eq:BW}
    \rm BW = 1.5 \times \frac{2 \times IQR}{N^{1/3}},
\end{equation}
where BW is the bandwidth, IQR is the interquartile range\footnote{Defined as the distance between the 75\% and 25\% quartiles.} and N is the number of points in the slice. Equation \ref{eq:BW} is 1.5 times the optimal bin width of a histogram characterized by a given IQR and N \citep{scott1979optimal}. The factor 1.5 is empirically defined to slightly reduce the noise while preserving global trends. For each distribution, we trace the kernel peak density. However, as we adopt the normalized kernel, the peak density is directly related to the ``width'' of the observed distribution. Namely, high peaks denote narrower distributions, whereas low density values are related to broader distributions. We use a bootstrap technique with N=1000 repetitions to assess errorbars. We define the Star Formation Main Sequence (SFMS) as the slice containing the best linear fit of the BC galaxies. This procedure (in gif format\footnote{That can be viewed by opening the PDF in Adobe Reader 9.}) and results are presented in Figs.~\ref{fig:gif_evolution} and \ref{fig:gif_evolution_faint} for cluster and field galaxies, respectively. For better visualization, we further add an inset with the variation of the Full Width Half Maximum (FWHM) in each panel to quantify the mixture of galaxy population properties in each slice. Each curve covers roughly the same variation in the SFR vs. $\rm M_{stellar}$ plane, namely from the beginning of the BC to the end of the RS. Nevertheless, since the comparison is done in the SFR vs. $\rm M_{stellar}$ plane, we are not directly addressing the mechanism navigating galaxies from one slice to another. Rather, we are comparing galaxy properties with a similar SFR and $\rm M_{stellar}$, but in different environments.

\begin{figure*}
    \centering
    \animategraphics[width =  0.8\textwidth,controls,loop]{3}{Figures/Gif_Cluster/Hist_Evolution_Smooth_Cluster_No_Mass_Separation_v2_}{0}{17}
     \caption{
     This image is originally a GIF. The image seen is the last frame/slice of the following procedure for cluster galaxies: 1) select galaxies within the solid black line slice highlighted in panel [a]; 2) build a probability density histogram for each property (Age, [Z/H] and TType). In this case, we use the Scott criteria \citep{scott1979optimal} to define the bin size in order to account for different number of points in each slice; 3) for each case we create a normalized epanechnikov kernel density estimate with bandwith set to 1.5 times the histogram bin; 4) we then estimate the peak density, FWHM and related errorbars using a bootstrap technique (with N = 1000 repetitions) for each slice. The curves in panels [b], [c] and [d] are the evolution of the peak density (large panel) and FWHM (miniature panel) as we progress from the top most slice to the bottom one. We color points and histograms according to the region in which the slice is (BC - blue, GV - green, RS - red). For completeness, as we evolve from top to bottom slice, we maintain the distributions from previous slices, which is shown as faded gray in each plot. The purple lines in panel [a] denote the slice containing the SFMS, which we then use the $\rm \Delta SFMS$ to report our results. In the same panel, we also add an arrow indicating increasing $\rm \Delta SFMS$. Finally, in panel [d] we stress through a black arrow the significant TType transition experienced by galaxies during the GV. Due to the use of a normalized kernel, there is a relation between peak density and FWHM, namely increasing peak density means decreasing FWHM.}
    \label{fig:gif_evolution}
\end{figure*}

We next provide an overall description of the observed trends for galaxies in clusters. We quantify the trends using the slice's offset from the SFMS (shown in purple in panel [a]), namely $\rm \Delta SFMS$, which is defined as the perpendicular distance with respect to the slices' slope (see black arrow in panel [a]). At the beginning of the BC ($\rm \Delta SFMS = 0.5)$, the age distribution (panel [b]) is highly peaked at $\sim 1.5$ Gyr, due to the active star formation in these systems. For the [Z/H] (panel [c]), we find the opposite trend, namely the most star-forming BC galaxies have a broad [Z/H] distribution peaking at low stellar metallicity ($\sim -0.5$). LTGs dominate galaxy morphology distribution at $\rm \Delta SFMS = 0.5$. The evolution towards $\rm \Delta SFMS = 0.33$ (the last slice before GV) is then characterized by: 1) age distributions becoming broader and peaking at older values; 2) [Z/H] distribution continues to be broad and evolve almost in a coeval way towards higher stellar metallicity; 3) there is a decrease in the peak value for TType distribution, but still characteristic of LTGs (TType > 0). The evolution in the GV ($\rm -0.5 < \Delta SFMS < -0.83$) indicate several processes happening simultaneously in galaxies. We find that age distribution have an almost constant peak age, but with increasing FWHM, which suggest a mixed population with different recent star formation history. Regarding [Z/H], the GV is roughly the region where distributions start to become narrower and with an increasing peak. However, we find the more striking result in morphology, for which the peak TType evolves from greater to smaller than zero in a single slice ($\rm \Delta SFMS$ from -0.67 to -0.83), despite a large FWHM and evidently bimodal TType distribution. From the GV on, age distributions have higher FWHM and an increasing peak age until the end of the RS ($\rm \Delta SFMS \leq -2$), in which there is a decrease in the FWHM. In panel [c], galaxies evolve towards an asymptotic value as they progresses in the RS, which is characterized by increasingly narrower distributions peaking at $\rm [Z/H] \sim 0.1$. We find similar trends for morphology in the RS too. There is an increasingly excess of ETGs as we progresses towards the RS tail, as expected. In particular, at $\rm \Delta SFMS = -1.33$ most of the galaxies are ETGs, despite not yet being completely quenched. This suggests that galaxies may reach their final morphological stage before the full star formation quenching.

In Fig.~\ref{fig:gif_evolution_faint} we show the result for field galaxies. We find similar evolution for the cluster and field galaxies evolution across the SFR vs $\rm M_{stellar}$ plane, which indicates that their difference is more related to the acting quenching mechanisms instead of the galaxy property itself. The GV also denotes the region in which we find the transition towards broader and narrower age and [Z/H] distributions, respectively, and a significant TType peak variation from LTG to ETG morphology. However, we stress that the difference we find for age distributions at the end of the RS ($\rm \Delta SFMS = -2.33$) is broader than what we find for clusters. This possibly highlight the clusters influence to quickly evolve galaxies towards higher ages (which means lack of star formation), whereas field galaxies may host more recent star formation episodes, resulting in broader distributions for field galaxies age at the end of the RS in comparison to clusters.

\begin{figure*}
    \centering
    \animategraphics[width =  0.8\textwidth,controls,loop]{3}{Figures/Gif_Field/Hist_Evolution_Smooth_Field_No_Mass_Separation_v2_}{0}{17}
     \caption{The same as Fig.~\ref{fig:gif_evolution}, but for field galaxies.}
    \label{fig:gif_evolution_faint}
\end{figure*}

Using the SFR vs $\rm M_{stellar}$ diagram, we dissect how galaxy properties evolve as they go from the BC to the RS. An important feature is the plateau followed by an increase we find in age for galaxies in clusters. It is characterized by a low peak density, which means that each region composing the plateau has a high FWHM, thus indicating a mixture of galaxies with several ages in a given slice. The more striking result the broader distributions, but with increasing peak ages as we approach the RS tail. In other words, moving from one slice to the following one just moves the entire distribution towards higher ages. This is a first suggestion on how the cluster environment feeds the GV region in a continuous form due to its infall rate. Galaxies seems then to evolve in a coeval manner from the GV to the RS, which results in the plateau of equal peak density and increasing age. Moreover, ``the end of the line'' is the tail of the RS, for which the cluster environment accumulates quenched galaxies. With respect to [Z/H], our results suggests an upper achievable limit due to the lack of further star formation events when reaching the RS. At last, galaxy morphology changes towards early-type shapes in just a few ``slices'' for both field and cluster environment, which suggests that most of the morphological transformation (if it happens) takes place in the green valley region. Finally, the plateau in age occurs exactly when we find peak densities towards lower values of TType.

\subsection{Dependence on Galaxy Stellar Mass}

\begin{figure*}
    \centering
    \includegraphics[width = \textwidth]{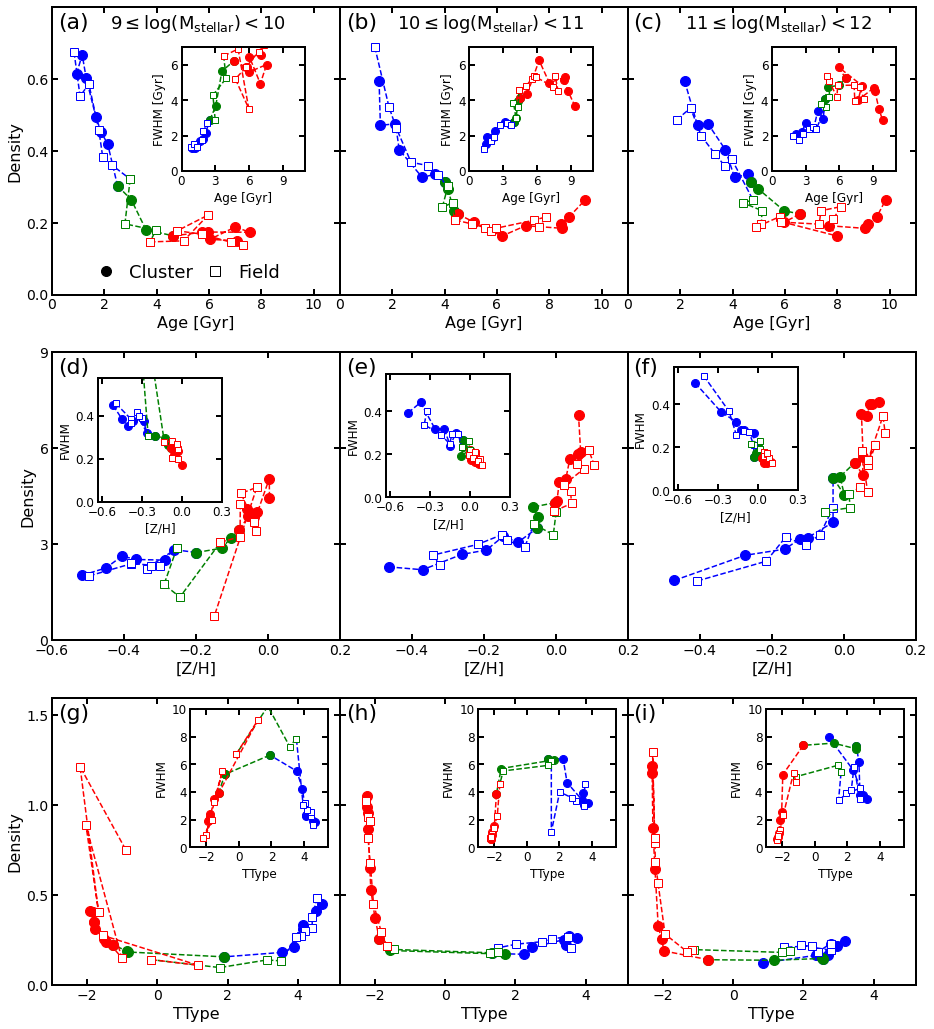}
    \caption{The resulting curves of the same procedure shown in Figs.~\ref{fig:gif_evolution} and \ref{fig:gif_evolution_faint} for cluster (filled circle) and field (white-faced squares) galaxies according to their stellar mass (each column). Colors denote if galaxies belong to BC (blue), GV (green) or RS (red). While the large panels display the peak of the normalized epanechnikov kernel density distribution as galaxies progresses from BC to RS, the inset in each panel shows the FWHM variation for each case.}
    \label{fig:hists_results}
\end{figure*}

Environmental mechanisms such as RPS and TML depend on both environment and galaxy halo mass. More massive galaxies are expected to be less prone to environmental effects of cluster environments. Here we use galaxy stellar mass as a proxy for full galaxy mass. In this section, in addition to the field-cluster separation, we separate galaxies according to stellar mass bins, similar to  Fig.~\ref{fig:Cluster_vs_Field_properties}. We then follow the same procedure as Figs.~\ref{fig:gif_evolution} and \ref{fig:gif_evolution_faint}. The results are presented in Fig.~\ref{fig:hists_results}. The stellar mass bins are chosen to guarantee robust statistics in every slice, while probing a sufficient variation to observe possible galaxy stellar mass related effects. 

Our results suggest that galaxies in different stellar mass bins experience different galaxy property variations during the same transition in the SFR vs. $\rm M_{stellar}$ plane. In panels [a], [b] and [c], the BC galaxy age distribution peaks are increasing with increasing stellar mass. Namely, for low mass galaxies (panel [a]), the top BC (first ``slice'') age distribution peaks at $\sim 1.5$ Gyr, while it peaks at $\sim 2$ and $\sim 3$ Gyr for intermediate (panel [b]) and high (panel [c]) mass galaxies. We find similar trends for [Z/H]. Panels [d], [e] and [f] indicate an increasing peak density of [Z/H] with increasing stellar mass, which indicates an excess of galaxies with higher metallicity ([Z/H] > 0.0) at the end of the RS (``last slice''). With respect to morphology (panels [g], [h] and [i]), the low mass bin is the only case where the peak density in the BC is greater than those at the end of the RS. The peak density at the end of the RS increases towards greater masses. 

After describing differences with respect to stellar mass, we now focus in the comparison between cluster and field systems. Although similar for low mass galaxies (panel [a]), age distribution peaks for intermediate (panel [b]) and high (panel [c]) mass galaxies in clusters reach older ages at the end of the RS in comparison to their field counterparts, despite starting roughly at the same peak density in the beginning of the BC. On the other hand, [Z/H] differences in peak densities happens mostly in the beginning of the BC (at $\rm \Delta SFMS > 0$). Clusters galaxies have lower [Z/H] peak densities in comparison to the field in the first $\sim 1-2$ slices. However, they converge roughly to the same peak density as we approach the RS tail. Finally, we do not find significant differences regarding morphology (TType). This suggests that the differences between field and cluster galaxy properties are mostly relevant in massive systems, which probably are more efficient in using stellar feedback to form new stars when in low density environment, whereas the cluster environment removes the galaxy's gas and prevents further star formation. This results in the plateau (more mixed distribution) in age observed for the field at the end of the RS in comparison to cluster environment.

Again, we stress that, in this case, we are not tracing the environmental/internal quenching mechanism itself, but the variations on galaxy properties with respect to location in the $\rm M_{stellar}$ vs. SFR diagram. The trends we find for different stellar masses are in agreement with works indicating that more massive galaxies are less affected by their environment in comparison to low mass systems \citep[e.g.][T20]{2010ApJ...721..193P}, which is expected since more massive systems form and assemble their stellar population at earlier times. Interestingly, in panel [a] of Fig.~\ref{fig:hists_results} we do not find significant differences between cluster and field galaxy ages. This may indicate: 1) in the field, AGN and stellar feedback driven gas outflows are sufficient to quench low mass galaxies, which is in agreement with T20; 2) the similar start and end points we observe indicate the cluster environment causing a more rapid quenching, while low mass galaxies have a similar ``fate'' indistinctly of the operating mechanism. The more striking results concerning cluster vs. field systems are those for intermediate and high mass galaxies in the RS, for which we find more broader age distributions for the tail-end RS galaxies in the field in comparison to clusters. This may follow from more massive systems in the field being able to produce a small amount (small enough to be RS galaxy), that otherwise is shutdown when in clusters due to the presence of environmental quenching mechanisms. Additionally, an important result comes from the TType panels [g], [h] and [i], namely galaxies suffer a rapid morphological transition in a single ``step'' of the GV. This provides further insight on whether SFR or Morphology changes ``faster''. Thus, a pivotal tool to understand galaxy is evolution especially in clusters is the focus of the next section: the infall time.

\section{Environment at Work: Cumulative Quenching in Galaxies}

Environmental quenching include multiple mechanisms which are inherently non-linear and complex. A fundamental parameter to understand how long a galaxy has been experiencing the cluster environment, traced by the time since infall. An important difference between galaxies experiencing the cluster for the first time and long-time members are their orbits. This fact translates to a well defined trajectory in the Phase-Space (see Fig.~1 of \citealt{2017ApJ...843..128R}), which enables an estimate of the infall time from galaxy position in the Phase Space. However, observationally we are limited to the projected phase space (PPS), but combined with N-body and cosmological simulations we can relate the galaxy's location in the PPS and time since infall.

We build PPS diagrams following the prescription of \cite{2017ApJ...843..128R}. The x-axis correspond to the projected radial distance and y-axis the velocity along the line of sight, both with respect to the clustercentric coordinates. We normalize the x and y-axis by the cluster's virial radius and velocity dispersion, respectively, in order to enable comparisons between clusters with different properties and create a stacked version of the PPS. With respect to the relation between location in the PPS and infall time, we adopt the ``New Zones'' (PNZs, hereafter) presented in \cite{2019MNRAS.484.1702P}. In this case, time  since infall is defined as the time since galaxy first crossed the virial radius. They used the YZiCS simulation to define quadratic functions fitting the observed time since infall distribution in the PPS. However, their region 7 is quite narrow and thus we decided to join regions 6 and 7 into a single region characterized by the mean infall time of both regions (see their Fig.~4 and Table 1). The PNZs are defined within one virial radius, thus the following analysis is limited to this radial limit. Still, a word of caution is needed regarding the way the PPS is discretized. Backsplash galaxies -- galaxies that have already completed their first pericentric passage and were thrown back to the cluster outskirts -- may influence our results when we examine the PPS since this population may suffer partial quenching in comparison to those first infalling. Even with cosmological simulations, the location in the PPS dominated by backsplash galaxies is not well defined. For instance, backsplash galaxies are mainly found in the $\rm [1 < R/R_{200} < 1.5] \times [0 < |V_{LOS}|/\sigma_{LOS} < 1]$ region, but even in this region they only account for 30\% of the whole galaxy population \citep{2011MNRAS.416.2882M}. An important feature of our analysis is the limitation to $R/R_{200} = 1$, which we expect to minimize the effect of backsplash galaxies. \cite{2021MNRAS.503.3065S} conclude that backsplash galaxy property distributions do not strongly affect the PPS based results.

\subsection{Galaxy Properties as a Function of Infall Time}
\label{sec:Galaxy_Properties_Infall_Time}
\begin{figure*}
    \centering
    \includegraphics[width = \textwidth]{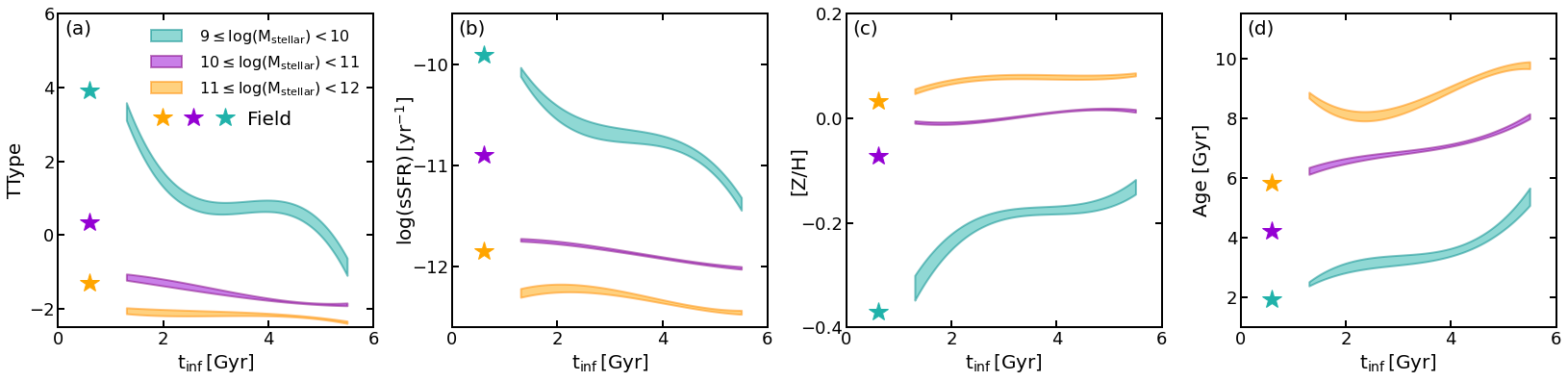}
    \caption{Relation between galaxy properties and time since infall. We present the relations for TType (left), sSFR (center-left), [Z/H] (center-right) and age (right). We divide galaxies according to their stellar mass (cyan, purple and orange colors). See panel [a] for the stellar mass intervals.}
    \label{fig:infall_time_relations}
\end{figure*}

We use the mean infall time of each PNZ to establish how galaxy properties evolve when infalling into clusters. We separate galaxies according to the same stellar mass bins of Fig.~\ref{fig:Cluster_vs_Field_properties}, but here the analysis is limited to galaxies within one virial radius. We compute the galaxy property median value in each of the 7 PNZs. To estimate the associated variance, we use a bootstrap technique: 1) we select N values of the observed distribution in a random way and with replacement; 2) We calculate the variance using the interquartile range as $\rm Q_{\sigma} \sim 0.7407 \times IQR$; 3) we repeat this procedure 1,000 times; 4) we then define the variance as the median of the $\rm Q_{\sigma}$ distribution. We present the results for TType, sSFR, [Z/H] and age in Fig.~\ref{fig:infall_time_relations}. For comparison, we add the median values of the field sample as colored stars. As a first approximation, we quantify variations and differences via a linear fit. However, this is only a first approximation and does not take into account the full behavior of the relations. We present the results for the linear fits in Table \ref{tab:linear_fit}. First, we notice how field galaxies are younger, more star-forming, more metal-poor and higher TTypes (LTGs) in comparison to those in clusters, irrespective of stellar mass. Moreover, differences are larger with increasing stellar mass. 

Next we detail the observed trends in Fig.~\ref{fig:infall_time_relations}. With respect to TType (panel [a]), the more massive galaxies (purple and orange) have early-type morphologies, irregardless of the time since infall. This trend is even more evident for galaxies with higher stellar mass, which at $\rm t_{inf} \sim 1.42$ Gyr have TType<-2. On the other hand, less massive galaxies (cyan) show a quick transition from late-type to early-type morphology. Namely, we find a $\rm \Delta TType \sim 3$ in $\rm \Delta t_{inf} \sim 1 Gyr$. In panel [b], we find similar trends for sSFR, which is almost constant for more massive galaxies and indicate more significant variations for the low mass regime. Furthermore, regarding low mass systems, the variations in TType and sSFR as a function of the time since infall, have a steeper slope (see Table \ref{tab:linear_fit}) for the former one. This is in agreement with galaxies reaching early-type morphologies before being full quenched. We focus exclusively on this topic in Section \ref{sec:ttype_ssfr}. With respect to [Z/H] (panel [c]), more massive galaxies have little variation with time since infall, suggesting these galaxies already reach their upper metal enhancement limit and lack gas for further star formation events. On the other hand, an important result is the trend we find for low mass galaxies. Namely, low mass galaxies show an increase in metalrichness at times of of $\sim ~1.5-2.5$ Gyr since infall and then level off to a constant value. This may indicate that galaxies quickly lose their gas component due to environmental effects, which prevents further enhancement in metallicity and results in the observed plateau at lower [Z/H]. Yet, the results in panel [d] are a consequence of the Age-Metallicity relation, in which the increasing [Z/H] means increasing age.

\begin{table}
\caption{Resulting slope and intercept of a linear fit for each curve in Fig.~\ref{fig:infall_time_relations}.}
\label{tab:linear_fit}
\resizebox{\columnwidth}{!}{%
\begin{tabular}{c|c|c|c|c|c}
\hline
$\rm X = log(M_{stellar})$ & y = ax + b & TType & \begin{tabular}[c]{@{}c@{}}log(sSFR)\\ {[}$\rm yr^{-1}${]}\end{tabular} & {[}Z/H{]} & \begin{tabular}[c]{@{}c@{}}Age\\ {[}Gyr{]}\end{tabular} \\ \hline
\multirow{2}{*}{$\rm 9 \leq X < 10$} & a & $-0.60 \pm 0.12$ & $-0.26 \pm 0.03$ & $0.032 \pm 0.011$ & $0.45 \pm 0.08$ \\ \cdashline{2-6}
 & b & $2.98 \pm 0.47$ & $-9.83 \pm 0.11$ & $-0.281 \pm 0.036$ & $1.96 \pm 0.30$ \\ \hline
\multirow{2}{*}{$\rm 10 \leq X < 11$} & a & $-0.17 \pm 0.03$ & $-0.07 \pm 0.01$ & $0.007 \pm 0.001$ & $0.40 \pm 0.05$ \\ \cdashline{2-6}
 & b & $-1.01 \pm 0.14$ & $-11.62 \pm 0.02$ & $-0.02 \pm 0.004$ & $5.57 \pm 0.21$ \\ \hline
\multirow{2}{*}{$\rm 11 \leq X < 12$} & a & $-0.09 \pm 0.03$ & $-0.07 \pm 0.02$ & $0.002 \pm 0.002$ & $044 \pm 0.08$ \\ \cdashline{2-6}
 & b & $-1.82 \pm 0.13$ & $-12.07 \pm 0.07$ & $0.072 \pm 0.008$ & $7.15 \pm 0.33$ \\ \hline
\end{tabular}%
}
\end{table}

The evolution of galaxy properties with time since infall depends strongly on galaxy stellar mass. Less massive systems are more affected by environmental effects, namely we find, for the same infall time interval, increasing variations with decreasing stellar mass. We investigate via a linear fit (as first approximation) that galaxies with different stellar mass reach the cluster environment with significantly different properties. This can be seen by the differences in the intercepts shown in Table \ref{tab:linear_fit}, in which all the intercepts are, at least, more than $1\sigma$ different between galaxy population with different stellar mass. We find evidence of galaxies reaching an asymptotic value in [Z/H], which decreases with decreasing stellar mass. Additionally, an important feature is the time it takes for the galaxy properties to reach the asymptotic value. We find that more massive galaxies reach more rapidly a constant value than low mass systems. This is directly related to how environment affect galaxy star formation. A quick halt of star formation disables further metal enrichment for the stellar component, resulting in an asymptotic value. The differences between the asymptotic value for different stellar masses then may be related to the amount of gas available for removal in these systems. Therefore, our results show how morphological and sSFR transitions happens at a different pace. Exploring the slope of the relations, we find that changes in TType are always steeper in comparison to sSFR. However, the difference depends on the galaxy's stellar mass too. This is an important piece of evidence on whether morphology or sSFR changes ``faster'' when galaxies infall in cluster environments, as discussed in the next section.

\subsection{Addressing Variations in Morphology and Star Formation Rate}
\label{sec:ttype_ssfr}
\begin{figure}
    \centering
    \includegraphics[width = \columnwidth]{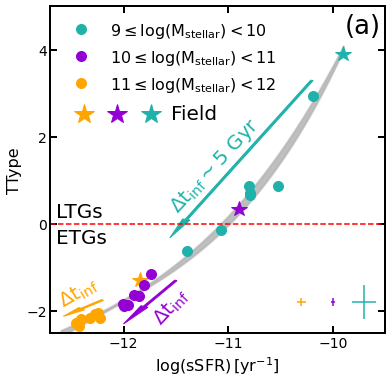}
    \caption{Relation between TType and sSFR for a fixed interval of time since infall (~5 Gyr). Each point represents the median TType and sSFR for a single PNZ (and hence infall time) in the PPS. In other words, we compress the results regarding TType and sSFR shown in Fig.~\ref{fig:infall_time_relations} in a single space, since they roughly cover the same time since infall range. We divide galaxies according to stellar mass (cyan, purple and orange). We present the mean errors in each case in the lower left corner. The colored arrows illustrate increasing infall time for each stellar mass bin. We add a spline fit and 1$\sigma$ error as a grey dashed area. The median values of the field counterpart is displayed as the colored stars and the separation between LTGs and ETGs is denoted by the red dashed horizontal line.}
    \label{fig:ssfr_vs_ttype}
\end{figure}

In this section we focus strictly on clusters to address the question of whether infalling galaxies reach their ``final stage'' in morphology or sSFR first? To do so we use the relations of TType and sSFR as a function of infall time. This way we can trace variations in TType and sSFR for a fixed time interval. We divide galaxies just as in Section \ref{sec:Galaxy_Properties_Infall_Time} to probe both stellar mass driven and environmental effects. We present the results in Fig.~\ref{fig:ssfr_vs_ttype}. Each point of each color corresponds to a single PNZ and, consequently, to a given time since infall. Colors represent the same stellar mass bins as Fig.~\ref{fig:infall_time_relations} (see legend in Fig.~\ref{fig:ssfr_vs_ttype}). At the bottom right we show the associated mean errors. We also include an illustrative arrow indicating the increase of infall time according to our results. The red dashed line represents the separation between LTGs and ETGS. Lastly, the grey dashed area and its width are the result of a spline fit and the 1$\sigma$ error, respectively.

As an overall trend, more massive galaxies first infalling in the cluster environment have TType smaller than zero and low sSFR. We find decreasing overall variations in both sSFR and TType with increasing stellar mass. A linear fit in the TType vs. sSFR relations results in slopes of  $2.74 \pm 0.36$ and $1.51 \pm 0.43$ for the high (orange) and intermediate (purple) stellar mass galaxies. In the former case, galaxies already enter the cluster as ETGs and then environmental effects (excluding mergers) do not cause significant variations in either TType or sSFR. Fig.~\ref{fig:ssfr_vs_ttype} results show unequivocally a higher variation in both TType and sSFR for less massive galaxies, which are more affected by the environment. Galaxies appear to have an especially rapid change in their morphology. However, comparing morphology and sSFR is not straightforward. We then make use of the fixed $\rm \Delta t_{inf}$ to detail these trends. Quantitatively, in a $\rm \Delta t_{inf} \sim 1$ Gyr, low mass (cyan) galaxies experience a variation in TType (sSFR) corresponding to $\sim 60\%$ ($\sim 28\%$) of the variation observed during the entire $\rm \Delta t_{inf} \sim 5$ Gyr. Finally, it is noticeable that low mass galaxies reach the TType = 0 before being completely quenched. Our results hence provide strong evidence of, with decreasing stellar mass, morphology rapidly changing due to the environment, rather than sSFR.

\section{Discussion}
\label{sec:Discussion}

\subsection{How Galaxies Evolve in Different Environments?}

A pivotal aspect of this work is exploring galaxy properties in the star formation vs stellar mass plane, which is widely used to understand galaxy evolution \citep{2010A&A...521L..53L,2011ApJ...742...96W,2013MNRAS.432.2488C}. By dividing galaxies into three stellar mass regimes\footnote{1) High Stellar Mass: $\rm 11 \leq log(M_{stellar}/M_{\odot}) < 12$; 2)Intermediate Stellar Mass: $\rm 10 \leq log(M_{stellar}/M_{\odot}) < 11$; and 3) Low Stellar Mass: $\rm 9 \leq log(M_{stellar}/M_{\odot}) < 10$}, we present in Fig.~\ref{fig:SFR_Mstellar_density} an increasing prominence of the star formation peak with decreasing stellar mass. This is in agreement with works suggesting a different star formation history for galaxies with different stellar mass \citep[e.g.][]{2010A&A...521L..53L,2017MNRAS.469.3670S}. Namely, more massive galaxies formed most of their stars at higher redshifts. Additionally, our results from Fig.~\ref{fig:sSFR_hist} indicate that most of the star formation in the local universe happens in field galaxies, i.e. galaxies in isolation. The higher fraction of star forming galaxies in the field compared to cluster at all stellar masses suggests that internal mechanisms have longer time-scales to fully quench star formation in galaxies. 

Exploring field and cluster galaxy properties in each region of the star formation rate vs stellar mass plane, we find increasing differences with increasing stellar mass. In particular, RS galaxies in the field are younger than those in clusters. With the age estimate adopted in this work, so our results indicate a more recent episode of star formation in the field population in comparison to its cluster counterpart. \cite{1999ApJ...527...54B} compare cluster and field galaxies at z$\sim$0.3 and showed the spectral differences between the two populations, especially in the D4000 break, favoring larger breaks for cluster galaxies. Additionally, early-type field galaxies in the local universe (z<0.044) are less massive and younger than their counterparts in clusters, despite similar spectroscopic characteristics \citep{2015ApJS..220....3K}. These three results are in agreement and points to early-type (mainly in the red sequence) field galaxies having a more recent episode of star formation in comparison to those in clusters. After reaching the red sequence, field galaxies are still able to reuse stellar feedback-enriched gas to form new stars, while in clusters environmental effects may remove this component preventing further star formation episodes. The comparison presented in panels [b], [e] and [h] of Fig.~\ref{fig:Cluster_vs_Field_properties} indicates that stellar metallicity does not differ significantly (especially in the red sequence) between galaxies in clusters or in the field. Despite the difference we find in age, galaxies seems to converge to $\sim$0.1 solar metallicity. This may indicate an upper limit of metal enrichment for galaxy stellar population, which has been proposed for galaxies in ``mini-halos'' \citep{2017MNRAS.465L..69C}. Also, the upper limit may be an artificial effect due to the SSP base used in the spectral fitting, which must be further investigated in a future work.

\subsection{Evolution of Galaxy Properties in High Density Environments}

By focusing on clusters, we detail how high density environments affect galaxy evolution. An important feature of galaxy quenching in clusters is the dependence on both host halo and  galaxy stellar mass. Following previous works, we present two conclusions prior to our discussion: 1) quenching of massive galaxies relies mostly on internal processes \citep{2010ApJ...721..193P}; and 2) just as environmental effects, gas outflows are increasingly relevant with decreasing stellar mass (T20). First, exploring Figs.~\ref{fig:gif_evolution} and \ref{fig:gif_evolution_faint} we detail the path towards the red sequence for field and cluster galaxies; and in Fig.~\ref{fig:hists_results} we do the same, but dividing galaxies according to stellar mass. We summarize our results as follows:
\begin{itemize}
    \item The normalized distribution for cluster galaxies' age is highly peaked at low ages during the blue cloud, which then evolves to a broader distribution during the green valley and in the beginning of the red sequence. During the red sequence phase, age is a broad distribution over a range of 5 Gyr and increasingly progresses to higher ages at each step in $\rm \Delta SFMS$ and then gets narrower and peaked at $\sim 10$ Gyr at the end of the red sequence;
    
    \item Metallicity and TType follow opposite trends in comparison to those found in age. Namely, the distributions in the blue cloud are broader and become narrower towards a single value as galaxies evolve towards the red sequence;
    
    \item In the field case, we find similar trends. However, in comparison to clusters, we find broader age distributions at the end of the red sequence. Also, the age at the end of red sequence peaks at a slightly lower value than in clusters ($\sim 9$ Gyr);
\end{itemize}
These results are related to the differences in how environmental mechanisms affect galaxies in comparison to those in the field. Following T20, the amount by which stellar metallicity is enhanced during quenching depends on the acting mechanism. Galaxies quenched via slow processes have time to increase their metallicity, while fast processes remove most of the remaining gas component in a short time-scale, preventing further enhancement in metallicity. This is in agreement with our results. The lower value of [Z/H] we find for low mass galaxies (Fig.~\ref{fig:infall_time_relations}) even after $\sim 5$ Gyr may be related to ram pressure stripping removing the galaxy's gas component, which is considerably more metal-poor than the massive counterpart. Regarding age, galaxies have a well known bimodal distribution. Investigating different slices of the star formation rate vs stellar mass plane, we find a constant peak value for age distributions after the green valley region.

We interpret the results above as follows: quenching results in a mixture of galaxies with different ages, seen as broader distributions; this mixture of galaxies half-way to the full quenching then evolves in a coeval way, for which we see displacement of the age distribution as we approach the bottom of the red sequence; at the end, galaxies have an upper limit in age, for which we note the cumulative effect forming the second peak of the distribution. However, a word of caution is needed due to the green valley definition. Although our results seems at first to be inconsistent with the current green valley, this apparent contradiction only reinforces how defining blue cloud, green valley and red sequence is far from trivial. Our results may indicate that these regions need more galaxy properties to be taken into account in order to provide a robust definition for a wide range of galaxy population and different redshifts. Besides, the commonly adopted definitions are not directly related to all the physical processes affecting galaxy evolution. Yet, our results need to be compared to different green valley definitions. Throughout our analysis we define the green valley following T20, which traces mostly recent star formation episodes. Another option is to a use more stellar population based definition, such as \cite{2019MNRAS.488L..99A} and Angthopo et al. (in prep.). For instance, galaxy property variations in different regions of the star formation rate vs. stellar mass plane depend on stellar mass. Exploring Fig.~\ref{fig:hists_results} we see higher variations in stellar population properties during the green valley for low mass galaxies. This increases the complexity of defining the green valley region for galaxies. If we want a definition of the green valley that clearly identifies galaxies undergoing a transition from star forming late-type to quenched early-type galaxies then we need a more sophisticated definition.  It is therefore important to ask what characterizes a green valley galaxy. We thus suggest the green valley galaxies to be characterized by: 1) mixed distributions of galaxy properties; 2) it is where we notice the morphological transition (if happening) from late to early type (Figs.~\ref{fig:gif_evolution} and \ref{fig:gif_evolution_faint}); 3) reduction in star formation at a given rate as a function of stellar mass; 4) increasing age and [Z/H].

\subsection{Morphological Transition Prior to Star Formation quenching?}

Whether morphology or star formation is first affected by the environment for an infalling galaxy has came to focus recently \citep[see][]{2009ApJ...707..250M,2019MNRAS.486..868K}. In this work we provide evidence of galaxies quickly changing morphology after entering the cluster, while their star formation rate keeps decreasing for a longer period. Dividing the star formation rate vs stellar mass diagram into slices of $\rm \Delta SFMS$, we find that, by the end of the green valley, the morphology distribution has a slight peak at early-type shapes, which then becomes extremely prominent as we approach the middle of the red sequence. This is trend we find at all stellar massess (panels [g], [h] and [i] of Fig. \ref{fig:hists_results}).

Furthermore, we use the projected phase space to address this question more directly. We relate the location in projected phase space with time since infall to study how morphology and specific star formation evolve. Our results provide further evidence that morphology is affected more quickly in comparison to specific star formation rate (Figs.~\ref{fig:infall_time_relations} and \ref{fig:ssfr_vs_ttype}). However, the time it takes for the morphological transitions depends on stellar mass. Most of the more massive galaxies already reach cluster environment with early-type morphologies, whereas the more drastic transition is seen for low mass systems. In particular, for low mass galaxies, the morphological transition happens in a time scale of $\sim$1 Gyr, while specific star formation decreases for $\sim$3 Gyr. After the transition towards ETG morphology, new quenching mechanisms can become important, such as morphological quenching. The stellar component evolution from disk to spheroidal shape may prevent gas density instabilities, which then prevents further star formation episodes.

\subsection{The Delayed-Then-Rapid Quenching Scenario}

In this section we interpret our results in the context of the Delayed-Then-Rapid Quenching Model \citep{2013MNRAS.432..336W}. According to this model, galaxies experience a delay phase after reaching the cluster environment, which takes around 1-4 Gyr, depending on the galaxy's mass \citep{2019A&A...621A.131M,2020ApJS..247...45R}. After the delay phase, galaxies are rapidly quenched mainly via a combination of environmental effects. \cite{2019ApJ...873...42R} indicate that ram pressure efficiently remove galaxy's gas component after galaxies reach a given ICM density. Prior to crossing the virial radius for the very first time, galaxies experience starvation and tidal mass loss. The former is identified as one of the main quenching mechanisms on long time-scales. \cite{2017ApJ...843..128R} suggest that, on long time-scales, galaxies lose a large fraction of their mass due to tidal mass loss. Our results indicate that galaxies have an almost constant age (panel [d] of Fig.~\ref{fig:infall_time_relations}) for $\rm \Delta t_{inf} \sim 2.5$ Gyr, which then increases in the last few Gyr. This  result is in agreement with the expectation of the delayed-then-rapid quenching model. The more striking results for high stellar mass galaxies is the evidence of less environmental effect due to the dependence of effects, such as Ram Pressure, on galaxy mass. In other words, we find an increasing environmental effect with decreasing stellar mass.

Starvation and tidal mass loss are processes that cause significant variations in star formation only on long time-scales. Our findings shows morphology varying more rapidly than specific star formation. Quantitatively, we find a variation from TType$\sim$3 to 0 in a time interval of 1 Gyr (Fig.~\ref{fig:infall_time_relations}, panel [a]), which then remains roughly constant. On the other hand, the plateau in specific star formation rate happens after $\rm t_{inf} \sim 3$ Gyr. This indicates that galaxy reach their ``final state'' in morphology before they are fully quenched. Furthermore, the time interval for morphological transformation is comparable to the expected for the slow phase of the delayed-then-rapid quenching model. \cite{2019A&A...621A.131M} compare the observed fraction of star formation as a function of the clustercentric radius with the predictions from the Millenium simulation and constrain the ``slow'' phase time-scale to be 1-2 Gyr. We interpret this as an evidence that, during the slow phase, galaxies are mostly morphologically affected, while the specific star formation rate keeps decreasing at a slow pace. After the slow phase, environmental effects remove the remaining gas, which quickly halt star formation of the, already early-type, galaxy. This process is increasingly relevant for decreasing stellar mass. It is important to stress that our claims contrast with those presented in \cite{2019MNRAS.486..868K}, which use the presence of late-type galaxies in the virialized region of the PPS to argue that morphology is changed only after the quenching of star formation. However, we highlight how galaxy properties strongly depends on redshift, their work focuses on systems at $\rm 0.4 \leq z \leq 0.8$ in comparison to the low redshift sample we adopt here.

\section{Conclusion and Summary}

In this paper we investigate the dependence of galaxy evolution on its host environment and provide a detailed view of the galaxy transition from the Blue Cloud to the Red Sequence. We use a sample of cluster member galaxies from an updated version of the Yang Catalog. In addition, we identify a field sample from the SDSS-DR7 database by finding all galaxies located at least 5 $\rm R_{200}$ away from any structure with $\rm log(M_{halo}/M_{\odot}) \geq 13$ listed in the Yang catalog. First, we recover a series of already known results regarding cluster vs. field galaxies and explore the mass dependence of such results. 

We then compare galaxy property distributions for galaxies in the Blue Cloud, Green Valley and Red Sequence according to their stellar mass and host environment. Red Sequence and Green Valley field galaxies hosted more recent episodes of star formation in comparison to their counterparts in clusters, for which we find an increasing relative difference with increasing stellar mass. These differences are then confirmed by dissecting the diagram in 17 different slices of $\rm \Delta SFMS$. This may indicate that massive systems in the green valley and red sequence are able to keep a small amount of star formation, otherwise halted in dense environments. We suggest that field galaxies are more prone to reuse their gas content, which would be removed in clusters, which also prevents further metal enhancement. Additionally, the greater difference is found in high mass galaxies, which appears to be more efficient in reusing stellar feedback. Our analysis provides additional observational evidence of a considerable fraction of galaxies undergoing a transition from early to late-type morphologies in the Green Valley. Yet, it is important to stress that galaxy evolution is not linear and galaxies can reform a disk after a merger \citep{2017MNRAS.465..547P}.

We directly assess how infalling galaxy properties vary as a function of infall time through the projected phase space. Galaxies with different stellar masses enter the cluster environment with different properties, for which we note an increasing environmental influence with decreasing stellar mass. We highlight the following results:
\begin{itemize}
    \item Low mass galaxies infalling in clusters in the local universe suffer a quick ($\sim$ 1 Gyr) morphological transition (LTG to ETG) just after crossing the virial radius for its very first time, while specific star formation keeps decreasing for a longer period of time ($\sim 3$ Gyr);
    \item Different stellar mass galaxies reach distinct plateau values in stellar metallicity, which are increasing for increasing stellar mass. 
\end{itemize}
Finally, in the context of the ``delayed-then-rapid'' quenching model, we suggest that the delay phase is mostly characterized by morphological transition, while star formation rate decreases at a slower pace. After, a combination of environmental effects (Ram Pressure Stripping, for example) further quench low mass galaxies by removing most of their gas content in a short time-scale ($\sim$ 1 - 3 Gyr). In particular, we interpret our results regarding stellar metallicity to follow directly from environmental quenching mechanisms removing all the gas of less massive, more metal-poor galaxies and disabling further metal enhancement via star formation in these systems.

\section*{Acknowledgements}

RRdC and VMS thank S. B. Rembold and T. F. Laganá for fruitful discussions on this topic. VMS acknowledges the CAPES scholarship through the grants 88887.508643/2020-00. RRdC aknowledges the financial support from FAPESP through the grant \#2014/11156-4. IF acknowledges support from the Spanish Ministry of Science, Innovation and Universities (MCIU), through grant PID2019-104788GB-I00. LCP thanks the National Science and Engineering Research Council of Canada for funding. This work was made possible thanks to many open-source software packages: AstroPy \citep{2018AJ....156..123A}, Matplotib \citep{2005ASPC..347...91B}, NumPy \citep{2011CSE....13b..22V}, Pandas \citep{mckinney2010data} and SciPy \citep{2020zndo....595738V}.

%%%%%%%%%%%%%%%%%%%%%%%%%%%%%%%%%%%%%%%%%%%%%%%%%%
\section*{Data Availability}
The data underlying this article will be shared on reasonable request to the corresponding author.



\bibliographystyle{mnras}
\bibliography{reference.bib} % if your bibtex file is called example.bib

%\begin{thebibliography}{99}
%\end{thebibliography}

%%%%%%%%%%%%%%%%%%%%%%%%%%%%%%%%%%%%%%%%%%%%%%%%%%

%%%%%%%%%%%%%%%%% APPENDICES %%%%%%%%%%%%%%%%%%%%%

\appendix

\section{Cluster Dynamical Property Estimates}
\label{AppendixA}
In this Appendix we briefly describe how virial mass, virial radius and velocity dispersion are estimated for each cluster. The velocity dispersion is estimated using the shiftgapper output final members list. Depending on the number of member galaxies, the cluster velocity dispersion estimate is then derived using gapper (N<15) or biweight (otherwise, \citealt{1990AJ....100...32B}) estimators, which is then corrected for velocity errors following \citep{1980A&A....82..322D} and resulting in a first estimate of the ``projected virial radius'' ($\rm R_{PV}$). This radius is then used to derive a first estimate of virial mass, \citep{1998ApJ...505...74G}:
\begin{equation}
    \rm M_{V} \sim \frac{3 \pi \sigma_{P}^{2} R_{PV}}{2G},
\end{equation}
where $\rm 3\pi / 2 $ is the deprojection factor and G is gravitational constant.
After, we apply a correction in the mass estimate due to the surface pressure term, for which we assume a Navarro-Frenk-White dark matter profile \citep[NFW]{1997ApJ...490..493N} with concentration given by
\begin{equation}
\rm    c = 4 \times \left(\frac{M}{M_{KBM}}\right)^{-0.102},
\end{equation}
where the slope and normalization are taken from \cite{2004A&A...416..853D} and \cite{2004ApJ...600..657K}, respectively. $R_{200}$ is then through the equation \citep{1997ApJ...478..462C}
\begin{equation}
    \rm R_{200} = \frac{\sqrt{3}\sigma_{P}}{10 H(z)}.
\end{equation}
Next, it is necessary to apply C-Correction \citep{1998ApJ...505...74G} in order debias the virial mass estimate due to the concentration of extended objects, such as clusters, which is given by
\begin{equation}
    \rm C = M_{V} 4\pi R_{PV}^{3} \frac{\rho (R_{PV})}{\int_{0}^{R_{PV}}} 4\pi r^{2} \rho dr \left[ \frac{\sigma_{P}(R_{PV})}{\sigma_{P}(r < R_{PV})} \right]^{2}
\end{equation}. 
Therefore, if $M_{V}$ is the virial mass after correcting for the surface pressure term in a volume of radius $R_{A}$, $R_{200}$ can be written as
\begin{equation}
    \rm R_{200} = R_{A} \left [  \frac{\rho_{V}}{200 \rho_{c}(z)} \right]^{1/2.4},
\end{equation}
where $\rm \rho_{V} = 3M_{V}/(4\pi R_{A}^{3})$ and $\rm \rho_{c}(z)$ is the critical density at a given redshift z. Finally, $\rm M_{200}$ is calculated by interpolating the the virial mass from $\rm R_{A}$ to $\rm R_{200}$. 

%%%%%%%%%%%%%%%%%%%%%%%%%%%%%%%%%%%%%%%%%%%%%%%%%%

% Don't change these lines
\bsp	% typesetting comment
\label{lastpage}
\end{document}